\documentclass[a4paper,12pt,notitlepage]{article}
\pdfoutput=1
\usepackage{helvet}

\usepackage{multirow}
\usepackage{rotating}
\usepackage{lscape}
\usepackage{appendix}
\usepackage{graphicx}
\usepackage[english]{babel}
\usepackage[section] {placeins}
\usepackage{mathtools}
\usepackage{amsmath}
\usepackage{gensymb}
\usepackage{tabularx}
\usepackage{textcomp}
\usepackage{float}
\usepackage{subfig}
\usepackage{listings}  
\usepackage{color}
\usepackage[nottoc,notlot,notlof]{tocbibind}
\usepackage{longtable}
\usepackage{soul} 

\usepackage{hyperref}
\hypersetup{
    colorlinks=true,
    allcolors= [rgb]{0.05,0.05,0.25},
}

\skip\footins 10pt plus 2pt minus 4pt 
\begin{document}
\vspace*{6\baselineskip}


\begin{figure}[!t]
\hskip -0.65cm
\includegraphics[width=17.9cm]{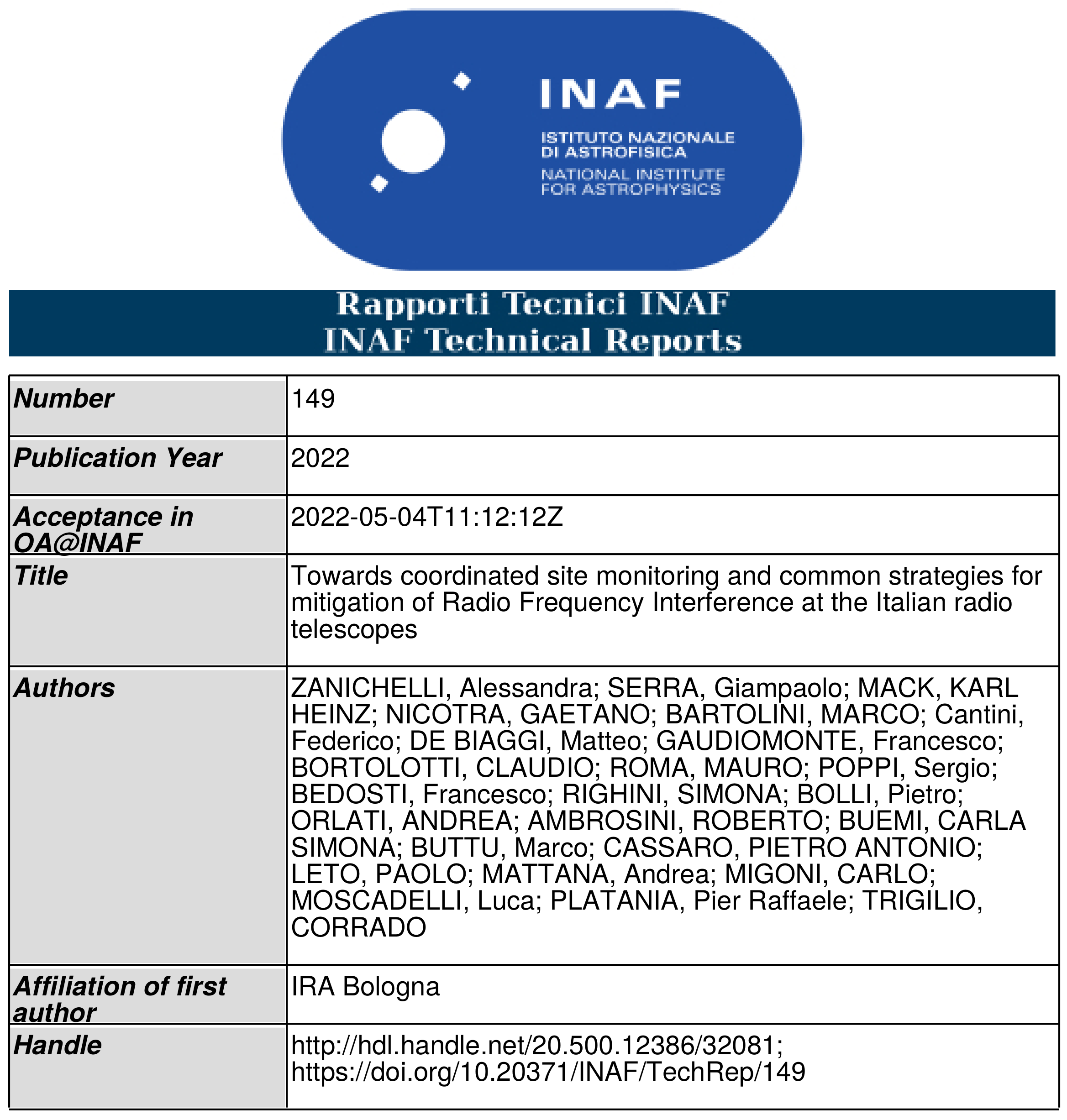}
\end{figure}
\pagenumbering{gobble}

\newpage
\pagenumbering{arabic}

\begin{center}
{\LARGE Towards coordinated site monitoring and common strategies for mitigation of Radio Frequency Interference at the Italian radio telescopes}

\bigskip
\bigskip

A. Zanichelli$^{1}$, G. Serra$^{2}$, K.-H. Mack$^{1}$, G. Nicotra$^{1}$, M. Bartolini$^{1}$, F. Cantini$^{3}$, M. De Biaggi$^{1}$, F. Gaudiomonte$^{4}$, C. Bortolotti$^{1}$, M. Roma$^{1}$, S. Poppi$^{4}$, F. Bedosti$^{1}$, S. Righini$^{1}$, P. Bolli$^{5}$, A. Orlati$^{1}$, R. Ambrosini$^{1}$, C. Buemi$^{6}$, M. Buttu$^{4}$, P. Cassaro$^{1}$, P. Leto$^{6}$, A. Mattana$^{1}$, C. Migoni$^{4}$, L. Moscadelli$^{5}$, P. Platania$^{1}$, C. Trigilio$^{6}$

\bigskip
\bigskip
\bigskip
\bigskip
\bigskip
\bigskip
\bigskip
\bigskip
\bigskip

Refereed by: E. Carretti$^{1}$
\bigskip
\bigskip
\bigskip
\bigskip
\bigskip
\bigskip

$^{1}$ {\small INAF - Istituto di Radioastronomia, via Gobetti 101, I-40129 Bologna, Italy}\\
$^{2}$ {\small ASI - Agenzia Spaziale Italiana, Selargius, I-09047 Cagliari, Italy}\\
$^{3}$ {\small Lib4RI-Eawag, Swiss Federal Institute of Aquatic Science and Technology, Ueberlandstrasse 133, 8600 Duebendorf, Switzerland}\\
$^{4}$ {\small INAF - Osservatorio Astronomico di Cagliari, Selargius, I-09047 Cagliari, Italy}\\
$^{5}$ {\small INAF - Osservatorio Astrofisico di Arcetri, Largo Enrico Fermi 5, I-50125 Firenze, Italy}\\
$^{6}$ {\small INAF - Osservatorio Astronomico di Catania, via S.Sofia 78, I-95123 Catania, Italy}\\

\end{center}

\newpage

\noindent
{\bf List of acronyms}
\begin{table}[ht]
\begin{tabular}{ll}
CASPER & Collaboration for Astronomy Signal Processing and Electronics Research \\
CRAF & Committee on Radio Astronomy Frequencies \\
DRGH & Double Ridge Guide Horn \\
DW & Dish Washer \\
FITS & Flexible Image Transport System \\
FPGA & Field Programmable Gate Array \\
GUI & Graphical User Interface \\
HiperLAN & High Performance Radio LAN \\
HPBW & Half Power Beam Width \\
iADC & Interconnect Analog to Digital Conversion \\
INAF & Italian National Institute for Astrophysics \\
ITU & International Telecommunication Union \\
LAN & Local Area Network \\
LEO & Low Earth Orbit \\
LPA & Log Periodic dipole Antenna \\
LTE &  Long-Term Evolution \\
NF & Noise Figure \\
NSM & Noise Signal Measured \\
OpenMP & Open Multi-Processing \\
PRIN & Progetti di Rilevante Interesse nazionale \\
RAS & Radio Astronomy Service \\
RBW & Resolution Band Width \\
RFI & Radio Frequency Interference \\
ROACH & Reconfigurable Open Architecture Computing Hardware \\
RX & Receiver \\
SARDARA & SArdinia Roach2-based Digital Architecture for Radio Astronomy \\
SR & Space Research \\
SRT & Sardinia Radio Telescope \\
TCP/IP & Transmission Control Protocol/Internet Protocol \\
TETRA & Terrestrial Trunked Radio \\
UC & Under Construction \\
UDP & User Datagram Protocol \\
UHF & Ultra High Frequency \\
VLBI & Very-long-baseline interferometry \\
WBLGB & WideBand Lowpass GigaBit \\
\end{tabular}
\end{table}
\newpage

\renewcommand*\contentsname{Table of contents}
\tableofcontents
\newpage

\begin{abstract}

We present a project to implement a national common strategy for the  mitigation of the steadily
deteriorating Radio Frequency Interference (RFI) situation at the Italian radio telescopes.
The project involves the Medicina, Noto, and Sardinia dish antennas and comprised the definition of a coordinated plan for site monitoring as well as the implementation of state-of-the-art hardware and software tools for RFI mitigation.

Coordinated monitoring of frequency bands up to 40 GHz has been performed by means of continuous observations and
dedicated measurement campaigns with fixed stations and mobile laboratories. Measurements were executed on the frequency bands allocated to the radio astronomy and space research service for shared or exclusive use and on the wider ones employed by the current and under-development receivers at the telescopes.
Results of the monitoring campaigns provide a reference scenario useful to evaluate the evolution of the interference situation at the telescopes sites and a case series to test and improve the hardware and software tools we conceived to counteract radio frequency interference.

We developed a multi-purpose digital backend for high spectral and time resolution observations over large bandwidths. Observational results demonstrate that the spectrometer robustness and sensitivity enable the efficient detection and analysis of interfering signals in radio astronomical data.
A prototype off-line software tool for interference detection and flagging has been also implemented. This package is capable to handle the huge amount of data delivered by the most modern instrumentation on board of the Italian radio telecsopes, like dense focal plane arrays, and its modularity easen the integration of new algorithms and the re-usability in different contexts or telescopes. 
\noindent

\bigskip
\bigskip
\bigskip{}
\bigskip{}
\bigskip{}

\noindent
\textbf{IMPORTANT NOTICE:} This document is primarily intended to provide an overview for radio-astronomical purposes only. Every other use of the collected data and/or considerations here reported is not in compliance with the authors wishes and therefore is in no way permitted. Characteristics like band occupancies or the presence of signals, even outside of the frequency bands allocated to the radio-astronomical service, have been observed with the only purpose of optimizing the design of the receivers that will be operating in that specific electromagnetic spectrum environment and/or the choice of the most appropriate observational strategy.

\end{abstract}
\newpage

\section{Introduction}

One of the most pressing problems for observational radio astronomy world-wide is the ever-deteriorating situation of Radio Frequency Interference (RFI).
Dynamical spectrum access is being increasingly exploited by transmitting devices, with new telecommunication technologies like mobile services and ultra-wide band devices more and more affecting operations at most of the radio observatories. 
The variety of man-made radio signals propagating in the environment surrounding the telescope sites, translates more and more frequently in RFI signals showing a complex behaviour in time and frequency that can hardly be modeled and corrected. Such a situation might pollute the radio astronomy receiver frequency bands and cause an irreparable data loss.

As sources of interference continue to proliferate, the requirements of astronomical observations become increasingly demanding in terms of frequency range, bandwidth and sensitivity, but also in the capability of RFI monitoring and mitigation.
Nowadays, the frequency bands assigned to passive science by art. 29 of the International Telecommunication Union (ITU) Radio Regulations, although still fundamental for radio astronomy, are no longer sufficient to perform cutting-edge science \cite{craf05}.
Wide-band receivers observe well beyond the limits of the passive bands primarily assigned to the Radio Astronomy Service (RAS), as also acknowledged by the national frequency allocation plan \cite{pnrf}. The increased sensitivity of modern, cryogenically cooled receivers and the availability of large aperture dishes imply that very low levels of RFI signal intercepted by the antenna beam pattern are considered detrimental for radio astronomy data acquisition, as defined in terms of the system noise (ITU-Recommendation RA.769). Moreover differences between interferometers and single-dish instruments must be accounted for in the treatment of RFI. While very-long-baseline interferometry benefits from the intrinsic decorrelation of signals acquired in different environments \cite{baan10}, single-dish radio telescopes are particularly vulnerable to RFI, as the astronomical and the disturbing signals are coherently added.

The avoidance strategy by itself can no more represent an adequate solution to the RFI problem, which can be better addressed with a more structured approach. A variety of methods should be implemented to address RFI mitigation, ranging from anticipatory actions to the excision or cancellation of RFI signals in the post-detection stage (see \cite{baan19} for a review). Preventing/proactive measures like monitoring campaigns are necessary to intervene on the local RFI environment on a regulatory level under local, national or regional authorities. Moreover, a continuous monitoring of the RFI scenario at the telescope sites is propedeutic to reactive measures, like the implementation of on-line and off-line RFI cleaning in the data. The same technological advances that fuel the proliferation of interference also produce technical solutions for RFI real-time, on-line excision. State-of-the-art integrated circuits are widely used in the fabrication of fast and reconfigurable backends for radio astronomy. They allow extremely high computation power, making it possible to identify on-line the frequency channels bearing unwanted signals, and to subsequently mitigate them by elaborating the signal even at the rate of the original sampling, prior to operating any data averaging. As a result, the loss of data can be reduced to a minimum. 
Complementary to on-line RFI mitigation, which poses strong requirements on the processing capabilities of the acquisition chain, off-line techniques are applied in the post-detection phase. Different strategies are adopted to identify and mitigate the many different types of RFI, including statistical methods capable to pinpoint RFI characteristic patterns in the time and frequency domains or threshold-based methods to detect more irregular RFI. Recent advances in deep learning techniques allowed the development of algorithms based on neural networks, that have been adopted both for single-dish and interferometric radio astronomical data.

In this report we present a similar multi-faceted approach that the Italian National Institute for Astrophysics (INAF) has started in the last years to deal with the increasing incidence of RFI at the three Italian radio astronomical sites.
The impact of RFI has recently greatly intensified at the telescope sites not only due to its abundance and appearance at progressively higher frequencies, but also because of the increasing usage of these antennas for single-dish wide-band observations.
A coordinated approach for the mitigation of RFI has started, comprising a national plan for the monitoring of the three observing sites and the development of on-line and off-line RFI mitigation tools. 
Preventive monitoring campaigns are being conducted to characterise the radio environment at the sites through well defined measurement procedures and the RFI spectral occupancy within the frequency bands of the radio astronomy receivers. This is accomplished by means of fixed and mobile instrumentation through well defined, common procedures. 
Hybrid approaches accounting for the different telescope sensitivity, the instrumentation used for RFI monitoring and the local orography pecularities have been experimented. In recent years, ROACH-based digital backends exploiting the publicly available software by the CASPER consortium have been installed at the Italian radio telescopes (e.g. \cite{melis18}). The experience gained in the use of such systems together with the acquired knowledge of the RFI scenario at the sites is motivating the development of in-house firmware dedicated to on-line RFI removal on board of digital spectropolarimeters. In parallel, the design of software tools specifically tailored for off-line RFI removal in single-dish data from the Italian radio telescopes is ongoing.

This report is structured as follows: in Section~\ref{project} an overall description of the project undertaken at the Italian radio telescopes for RFI mitigation is provided. The activities related to coordinated site monitoring, development of a digital spectrometer and of an RFI flagging tool are reported in Sections~\ref{monitoring}, \ref{spectrometer}, and \ref{dishwasher} respectively. Our conclusions are presented in Section~\ref{conclusions}.

\section{The Project}\label{project}

The three Italian radio telescopes - Medicina, Noto and Sardinia Radio Telescope (SRT) - are located in the Po Valley, in Sicily, and in Sardinia regions respectively. Most of the Italian territory is densely populated, and the aforementioned sites are forcibly surrounded, within a few tens of kilometers, by towns or cities. Medicina is the most exposed site; for this reason, it is in a front position to anticipate the evolution of the RFI situation at the other radio observatories in a near future. The orography of the SRT site is characterised by the presence of mountain hills surrounding the antenna, while the landscape at Noto is flatter except for some hills in the north-eastern direction. On the contrary, the Medicina telescopes are surrounded by a flat landscape. 
This results in a similar RFI scenario at the sardinian and sicilian sites. Such a complex scenario and the rapid evolution of the man-made impact on the radio astronomical observations demanded for a coordinated approach towards RFI mitigation in Italy.

The project undertaken at the Italian radio telescopes was
initially funded within the framework of the INAF Techno PRIN 2012 (project title ''RFI mitigation at the Italian radio telescopes'', P.I. Dr. Karl-Heinz Mack) and aims at creating re-active RFI mitigation procedures, which complement the various pro-active measures that must be put in place in a coordinated manner at the telescope sites. To reach this goal, three main activities have been identified.
In the first place, the knowledge of the local RFI environment is mandatory to identify and characterize the most dangerous disturbers of scientific observations and is propedeutic also to the implementation of further hardware/software mitigation strategies. To this aim RFI coordinated surveys have been realized in the frequency range of the receivers currently used for radio astronomical observations and in particular in the  narrower frequency bands allocated to RAS and Space Research (SR) acknowledged by the ITU and the national authorities.

The second activity is focused on the acquisition of skills in the in-house development of firmware for spectropolarimetric digital backends with high spectral and time resolution capabilities over large bandwidths. Building on previous experience on FPGA-based systems and the availability of such hardware at the three telescopes, the first prototype of a multi-purpose digital spectrometer has been developed based on ROACH boards.
As the three Italian antennas share the same telescope operating system \cite{orlati16}, the software implementation of the current prototype and future ROACH-based backends is easily exported from one telescope to another.
The common data formats delivered by the instrumentation on board of the Italian dishes allow for a coordinated approach also in off-line RFI mitigation, which is the subject of the third activity. Current and future multi-feed receivers on board of the Italian dishes require the design and implementation of software routines optimized to simultaneously handle multiple, big-sized data streams. This has been accomplished by developing a software tool for data inspection and flagging, specifically tailored to handle spectropolarimetric single-dish data output by the Italian radio telescopes. 

In the next Sections a description of the results achieved in the three activities discussed above is reported.

\section{RFI campaigns at the Italian Radio Observatories} \label{monitoring}

The RFI scenario at the Italian radio telescopes is continuously evolving with peculiarities typical of each local environment. The Medicina radio telescopes\footnote{http://www.med.ira.inaf.it}, the 32-m dish and the Northern Cross interferometer, are exposed to a wealth of artificial signals from the surrounding Po Valley region. The 32-m Noto\footnote{http://www.noto.ira.inaf.it} and the 64-m SRT \cite{Bolli, Prandoni} radio telescopes were built in South-Eastern Sicily and in Southern Sardinia respectively to define the much long as possible baseline for the Italian radio interferometer experiments. Moreover, the telescopes are located on different tectonic plates thus making possible to perform geodynamical studies.
Although these two sites were chosen after an accurate RFI investigation and the areas around the telescopes still remain sparsely populated, the RFI signals polluting the frequency bands of their receivers are steadily increasing.

Interfering signals can come from fixed transmitters located in the telescope surroundings, but also from very distant geosynchronous satellites or the ones orbiting around the Earth: all of them are seen by the telescope as sparsely distributed over time, frequency, amplitude, polarization
depending on the radio service type or on the accidental spurious signal characteristic.
Moreover, they typically are received from specific azimuth directions where inhabited areas, power lines, transmitting stations, satellites etc are located.
Even the accessory instrumentation and electronic devices in use at the observing site have to be properly designed to avoid self-induced disturbing electromagnetic signals. These would be detected by the telescope as invariant signals on the azimuth direction, that must be monitored and mitigated as well. 

RFI monitoring is conducted at the Italian radio telescope sites by measuring the presence of disturbing signals within the bandwidth of all the available receivers. In particular, the sub-bands allocated to the RAS and SR are extensively monitored. 
The total frequency range taken into consideration in the measurements activities presented here is 0.3 - 40 GHz. 
An upgrade of the telescope receiver suite up to 116 GHz is ongoing thanks to a National Operational Program funding assigned to INAF in 2019 by the Italian Ministry of University and Research \cite{Govoni}. In the near future, an extensive RFI monitoring will thus be needed to characterize the observing sites also at these higher frequencies.

\begin{figure}[h]
\centering
\includegraphics[width=24pc]{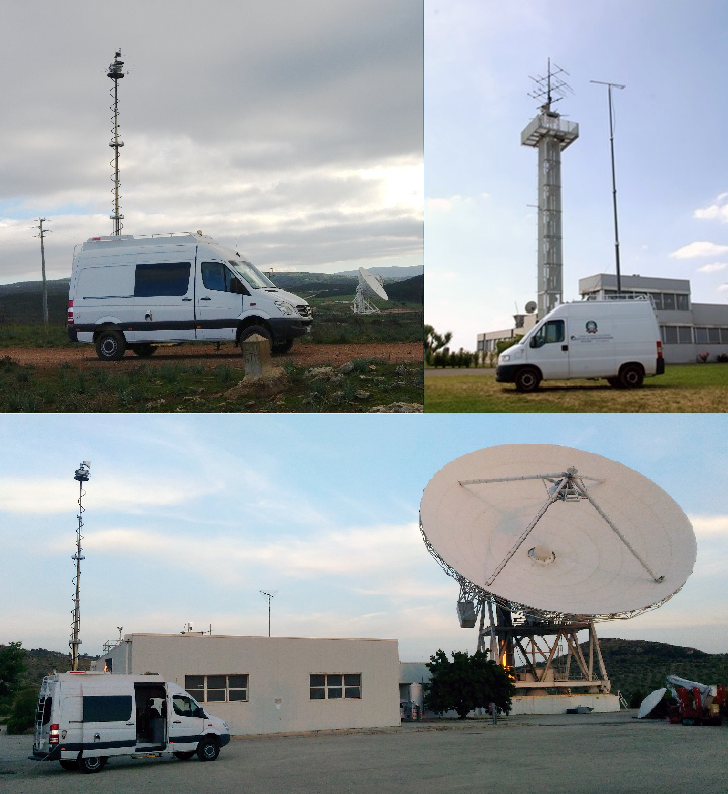}
\caption{SRT Mobile laboratory close to the SRT site (top left) and at the Noto radio telescope (bottom). Medicina mobile laboratory and 22-m-high tower in background (top right)}
\label{MobileLabs}
\end{figure}

The monitoring of the receiver frequency bands, especially those allocated to the RAS, has been routinely performed at the Medicina site since the beginning of the 1980s.  Such activities have been carried out by means of a fixed receiving station plus a mobile laboratory \cite{Ambros0, Ambros1, Ambros2, Ambros3, Monteb}, see Figure~\ref{MobileLabs}.
This longstanding experience was then transferred to the staff at the other two Italian radio telescopes, which have been equipped with instrumentation dedicated to the RFI monitoring as well (see Figure~\ref{MobileLabs}). At the SRT a mobile laboratory has been operating since 2009 with state-of-the-art instrumentation for RFI monitoring up to 40 GHz \cite{Bolli0, Bolli2}, see Figure~\ref{MobileLabs}. 
In addition, recently a new RFI fixed station started operating at SRT by exploiting the telescope itself and, hence, its high sensitivity as receiving antenna \cite{Serra2}.
A new portable receiving system working in the frequency range 0.3-18 GHz has been recently made available for RFI measurement at the Noto site \cite{nicotra}, as illustrated in Section~\ref{RFI Monitoring at Noto}.

The RFI groups at the three Italian radio astronomical sites work in strict collaboration, sharing instrumentation and measurement strategies as well as performing combined measurement campaigns to trace the evolution of the local RFI scenarios. At the same time the specific peculiarities of each site and telescope, like the local orography or the overall frequency coverage of the receivers, have motivated the development and use of different and complementary monitoring techniques.
\\
The information on the evolution of the RFI scenario (i.e. the characteristics of the signals polluting the receivers frequency band and their angular direction w.r.t. the telescope site) are usually shared with the telescope users, with the only purpose to minimize the astronomical data loss by giving them the opportunity to choose suitable countermeasures, if these latter are available at the telescope. For instance, such information allows the user to conveniently set the local oscillator frequency, down converting only the less-polluted portion of the selected receiver frequency band. Also, users can take advantage of other on-line and off-line strategies addressed to the mitigation or cancellation of the received RFI (see Sect.~\ref{spectrometer} and \ref{dishwasher}).

The RFI teams also represent the Italian radio astronomical community to both the national authority responsible for the spectrum regulatory issues (Ministry of Economic Development) and the Committee on Radio Astronomy Frequencies (CRAF), which acts on frequency issues for European radio astronomy and related sciences. Results from the monitoring campaigns are shared with these national and international authorities, reporting information on the RFI affecting radio astronomy observations and, in particular, the  RAS frequency sub-bands.   
In the framework of this national coordination, the following Subsections describe the results of measurement campaigns at the observing sites, highlighting the methods adopted, the RFI scenario and its time evolution, and some examples highlighting the importance of a cooperative effort among the RFI groups.

\subsection{RFI coordinated monitoring at the Medicina telescope}

Two radiotelescopes are operated at the Medicina station: the 32-m parabolic dish and the Northern Cross antenna array. The first one is a fully-steerable Cassegrain antenna offered to the radio astronomical community for single dish and VLBI experiments. The telescope is equipped with dual-polarization receivers distributed between the primary and the Cassegrain focal positions: four cryogenic (coaxial S- and X-band, C-low-band and dual-feed k-band) and two room-temperature ones (L- and C-high bands), operating in the frequency range 1.35-26.5 GHz.  A cryogenic, dual-polarization and dual-feed Ku receiver is under construction to fill the gap in the frequency range 13.5-18 GHz and this band is thus included in the RFI monitoring campaigns.
The Northern Cross is a transit interferometer composed by hundreds of parabolic dipoles arranged in a wide 'T' shape for a total collecting area of about 30000 square meters. The array observes in a frequency range of a few MHz centered at 408 MHz and is recently undergoing an upgrade of its northern arm receivers.

The fixed station operated by the local RFI group to monitor the RAS and the (wider) receiver frequency bands is equipped with several types of antennas covering the frequency ranges 0.1-12 GHz and 22-24 GHz. The antenna set (dual polarization: V+H) and the equipment to select, filter, amplify the signal coming from each antenna is mounted on a rotating mast located on the top of a 22-m high tower, equipped with an electro-mechanical rotor allowing a 360-degree monitoring in the azimuth plane. {\it Ad hoc} measurements are also performed by means of a mobile laboratory, a FIAT Ducato van equipped with a set of antennas, covering the frequency range .310-40 GHz, which can be mounted one at a time on a telescopic mast capable of a 360-degree rotation in the horizontal plane. Apart from the continuous monitoring routinely conducted to provide an up-to-date knowledge of the RFI scenario around the telescope site, further measurements are carried out whenever the telescope users report the detection of polluting signals during their scientific observations.  Tables~\ref{Medicina_FixedStationtable} and~\ref{MEDICINA_MObileStationTable} in Appendix A list the main characteristics of the Medicina fixed and mobile monitoring stations respectively.

The sistematic campaigns for RFI monitoring started many years ago have allowed to verify the long term,
ever-increasing level of man-made radio signals polluting the frequency band of the telescope receivers. As an example, in recent years the C-low and S-receiver frequency bands (1.5 GHz and 0.16 GHz wide respectively) have been polluted by signals from broad-band digital radio links, Wireless LAN and HiperLAN that are increasingly widespread in the populated areas around the Medicina site. Broad-band radio digital links are the main polluting sources even in the C-high and K frequency bands (1.2 GHz and 8.5 GHz wide respectively), even if at a lower level than the C-low- and S- ones, according to the reports from telescope users. RFI signals from the 5G mobile phone network have been recently detected in the K-frequency band, while polluting signals from 4G-LTE devices continue to occupy the L-frequency band. Furthermore, impulsive broad-band radar transmissions and satellite downlinks are detected in the L- and X- frequency bands respectively as well as at the frequencies of the under-construction Ku-band receiver, which is also polluted by emissions from digital links, short range devices and more recently by Starlink satellite signals.
Table~\ref{Med_RFIoccup} shows an estimate of the RFI spectral occupancy percentage at Medicina in the last years. These numbers cannot be directly compared with those obtained for SRT and Noto with a different {\it modus operandi} (see Section~\ref{RFI Monitoring Sardinia-Sicily}), but they however permit an evaluation of the local RFI scenario. The C-low receiver is by far the most polluted one, with an RFI spectral occupation percentage of 45\% and increasing in time. This is mainly due to disturbing signals from broadband digital links and HiperLAN emission. Broadband digital links are the main source of RFI also in the C-high and K frequency ranges, with a percentage occupancy equal to 25\% and 5\% respectively. Impulsive broadband radar transmissions and satellite downlinks are affecting the L frequency band, more significantly at the lower frequencies, and impacting at a much lower level also the X and Ku frequency ranges. The latter in particular is polluted also by emission from digital links and short-term devices.

\begin{figure}[h]
\centering
\includegraphics[width=38pc]{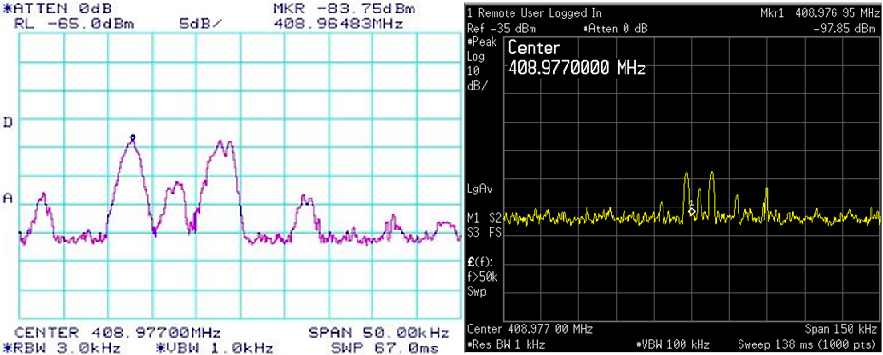}
\caption{RFI signal in the 406.1-410 MHz RAS primary use band, detected at Medicina with the RFI tower (spectrum on the left) and at the SRT site with the mobile laboratory (spectrum on the right). The RFI was due to a spurious emission in the signal transmitted by a LEO Spanish satellite called OPTOS n.39420}
\label{optos}
\end{figure}

This operational approach based on coordinated measurements with both the fixed and the mobile stations is efficient also in the detection of RFI affecting radio telescopes distant from each other. This is the case for instance, of the polluting signal emitted by a satellite and received in the 406.1-410 MHz frequency band (allocated to the RAS with primary use status) at both the Medicina and the SRT sites. Thanks to coordinated RFI monitoring, the source signal was identified and its frequency spectrum recorded by both the Medicina and SRT RFI stations, see Figure~\ref{optos}. 

\begin{table}[t]
\caption{Status, frequency bands and RFI spectral occupancy at the Medicina 32m antenna. The percentage in the last comlumn is the RFI spectral occupancy computed with respect to the receiver frequency band considering the worst case between the two linear polarizations}
\label{Med_RFIoccup}
\centering
\renewcommand\arraystretch{1.2}
\renewcommand{\tabcolsep}{0.5cm}
\begin{tabular}{l c c c}
\hline
 Receiver & Status & Frequency band (GHz) & RFI spect. Occupancy(\%) \\
\hline
L-low & working & 1.35-1.45  & 15 \\
L-high & working & 1.595-1.715 & 35 \\
S & working & 2.20-2.36 & 45 \\
C-low & working & 4.3-5.8 & 45 \\
C-high & working & 5.9-7.1 & 25 \\
X & working & 8.18-8.98 & 5 \\
K & working & 18.0-26.5  & 5 \\
Ku & under constr. & 13.5-18  & 15 \\
\hline
\end{tabular}
\end{table}

The spectra show the polluting signal received in a 50 kHz frequency range centered at 409 MHz after setting the spectrum analyzer with a spectral resolution of a few kHz in average mode. After accurate investigations, the signal turned out to be compatible with the signal carrier transmitted at 402 MHz by a Low Earth Orbit (LEO) satellite, known in the satellite catalogue as OPTOS n.39420. Medicina and SRT radio telescopes detected its emission during its transit at the Italian meridian. Such disturbing signal has been reported to the satellite company, who collaboratively acted on the down-link transmission to mitigate the interference. Since then, this spurious signal has been no more detected.

\subsection{RFI Monitoring Campaigns in Sardinia and Sicily with the SRT mobile laboratory}\label{RFI Monitoring Sardinia-Sicily}

In order to characterize the radio environment in the frequency bands at which the Sardinia and Noto telescopes operate, the local RFI teams put together resources and efforts to perform three intensive measurement campaigns in December 2014 and May 2016 at Noto and in November 2016 at SRT taking advantage of the coordination and funds from the INAF Techno PRIN project mentioned in Section~\ref{project}. Surveys were conducted in the frequency range 0.3-40 GHz by using, at both sites, the SRT mobile laboratory \cite{Bolli0, Bolli2}. Such measurements provided for the first time an overview of the RFI scenario surrounding the Noto radio telescope. At the same time, they allowed the update of the RFI scenario at the SRT just before offering the telescope for open-sky observations.
\\
At both sites the RFI teams carried out several radio measurements by choosing different configurations of the mobile station receiving system (see Appendix~\ref{RFIlab_conf_settings}) and using the following {\it modus operandi}. First, a suitable location on top of a hill close to the telescope is chosen to allow a free-space 360-degrees azimuth scan of the antenna mounted on top of the van mast. Secondly, the large frequency range is split into several sub-bands. For each sub-band, four azimuth scans (two for each polarization) are performed and the related spectra are acquired by the spectrum analyzer in max-hold mode by setting the instrument first in a broad-band (resolution bandwidth RBW $=$ 3 MHz), then in narrow-band (RBW $=$ 0.1 MHz) configuration. The 3 MHz- and 0.1 MHz-RBW spectra are used for broad-band/impulsive and narrow-band signal types respectively. In both cases, the operator sets a total number of spectral samples (up to a maximum of 8000 points allowed by the spectrum analyser on board of the SRT RFI mobile station) such that the resulting frequency bin is smaller than the chosen RBW. The spectrum analyzer RBWs were chosen after some preliminary measurements addressed to figure out how to better represent the RFI scenario surrounding the telescopes. At the top of a nearby hill the antenna is then lifted at a height of 10 m above the ground and a full horizontal plane (hereafter 'panoramic') spectrum is acquired every 360-deg rotation of the mast, lasting $\sim$60 seconds at a constant speed of $\sim$6 deg/s. In such a way four (H- and V- polarized, narrow- and broad-band configuration) spectra are available for each sub-band.  After completing all such scans the operator moves the van in a second strategic location, closer to the telescope and offering similar full view of the horizon, and a new acquisition is started with the spectrum analyzer in the same configuration as before. 
In such a way, homologous spectra of the same radio environment acquired at two different locations can be then compared. To this aim the operator initially selects and combines those spectra that better represent the RFI signals received in each of the telescope frequency bands. The channel gain (see Tables~\ref{SRT_MobileStation} and~\ref{Noto_MobileStation}) is removed from the data to retrieve the power level at the antenna input. Consequently, information on the received polluting signals like the maximum power level and its frequency, the bandwidth and polarization and the RFI type (impulsive, periodic, etc.) can be extracted.
Whenever RFI is detected in RAS subbands, signal maximization at three different measurement sites permits to identify the polluting source coordinates.
Comparison between levels of the same (steady) RFI signals in spectra acquired from different surveying sites would allow the identification of the effective RFI signal strength expected at the telescope location.
This {\it modus operandi} allows a wider characterisation of the RFI scenario, including those disturbing signals that are attenuated or not measured at all by the mobile station antenna at the location closer to the telescope, but that would however be detected by the telescope itself thanks to its greater height above the ground. In addition, it allows the identification of spurious signals generated locally by power lines, electronic apparatus and other active radio frequency sensors that, due to the site orography, would not be detectable by the mobile station at the location far from the telescope.

By means of the information collected in the measurement campaigns it has been possible to compute the lower limit for the RFI occupancy of the frequency bands for the existing and future receivers at the Noto and Sardinia telescopes (see Tables~\ref{SRT_MobileStation} and~\ref{Noto_MobileStation}). The uncertainty on the occupancy percentage is strictly related to the RF receiver sensitivity and  the spectrum analyzer settings of the mobile lab. To do that, for each frequency band we summed the spectral occupancy  of the signals classified as RFI on the basis of the comparison between homologous (same frequency range and polarisation) spectra. 0.1 MHz- and 3 MHz-RBW spectra are used to better characterize the power level and bandwidth of the RFI signals. Finally, the total RFI spectral occupancy is calculated in terms of percentage w.r.t. the extent of the receiver frequency bands and sub-bands.
\\
The radio monitoring reported in this Section lasted five consecutive days for each campaign. The 360-degrees panoramic spectra acquired during these measurements are characterised by an angular resolution equal to the antenna Half Power Beam Width. Given the duration of each panoramic scan, this means that the exposure time per pixel (i.e. the HPBW integration time) varies from a maximum of 8 s (for the LPA 370-10 antenna having a HPBW equal to 48 deg @ .370 GHz) to 1.1 s (for the DRGH 1840-A HPBW equal to 7 deg @ 33 GHz), see Tables~\ref{SRT_MobileStation} and~\ref{Noto_MobileStation} in Appendix B. The information provided by these panoramic spectra are thus to be considered as snapshot views of the instantaneous RFI landscape at the site. The database of panoramic radio spectra obtained during these campaigns does not give a complete representation of the local RFI scenario (which would be possible only by means of a continuous monitoring) and is referred specifically to the year 2016. Nevertheless, it provides a wealth of information which can orient the specific design of the ‘local optimum’ RFI mitigation strategy and is useful to prepare ad-hoc countermeasures. Also, this information can be compared with more recent measurements and thus used to evaluate the evolution of the RFI scenario.

In the following sub-sections~\ref{RFI Monitoring at SRT} and~\ref{RFI Monitoring at Noto}, we summarize the main results of the radio monitoring performed by means of the {\it modus operandi} described above. Moreover, we report some examples of further investigations triggered by the radio monitoring itself and aimed at identifying the direction and geographical position of RFI sources emitting in the RAS frequency bands. Finally, in Tables~\ref{SRT_RFIscenario} and~\ref{Noto_RFIscenario} we give an estimation of the RFI spectral occupancy percentage in the receiver frequency bands and sub-bands at SRT in Nov 2016 and at Noto in May 2016 respectively.

\subsubsection{RFI Monitoring Campaign at SRT}\label{RFI Monitoring at SRT}

The SRT is a general-purpose 64-m quasi-Gregorian reflector operated by INAF and the Italian Space Agency (ASI) for 80\% and 20\% of the available antenna time respectively. Currently SRT is equipped with four dual-polarization cryogenic receivers: a L- and P-band coaxial feed, a K-band 7-feed planar array and a C-high mono-feed operated by INAF and a single-polarization cryogenic X-band receiver dedicated to planetary probes tracking and radio science experiments and operated by ASI. Both single dish and VLBI observations are possible in the non-continuous frequency coverage 0.305-26.5 GHz \cite{Bolli3}. 

Since 2009 the SRT receivers frequency bands, especially those allocated to the RAS, have been regularly monitored mainly by means of the mobile laboratory. Before the start of SRT scientific observations in 2016 the local RFI team in collaboration with the Noto group intensified the RFI monitoring carrying out an intensive five-days measurement campaign in the frequency range 0.3-40 GHz. The setup of the mobile laboratory and its instrumentation during this campaign are summarized in Table~\ref{SRT_MobileStation}. Two measurement locations were chosen, the first one on a mountain hill, called Monte Ixi, located about 2 km from the telescope and the second on a pitch less than 1 km in the same valley where the telescope is placed. 
The main results of this RFI campaign are reported in Table~\ref{SRT_RFIscenario}. The first two columns list the SRT receiver names, their frequency ranges and the related sub-bands including the RAS and SR allocations, i.e. Exclusive Primary (EP), shared Primary (P) or Secondary (S) use status. The third column shows the noise floor level measured by the spectrum analyzer in each frequency range (representing the sum of the intrinsic system noise and the environment noise), corrected by the RF chain gain and expressed in terms of spectral power flux density ($dBW m^{-2} Hz^{-1}$). Considering homologous spectra, similar values of the noise floor have been measured in the 0.1 MHz- and 3 MHz-RBW spectra for all the frequency ranges, except for the P- and L-band ones. Due to broad-band electromagnetic disturbances radiated by power lines in the P band and by radar plants in the L band, at these frequencies the 3 MHz-RBW spectra, acquired at the location closer to the telescope, have a noise floor larger than that measured in the 0.1 MHz-RBW spectra at the same site. 
Since the power line polluting emission has a sporadic occurrence which depends mainly on weather conditions, in the third column of Table~\ref{SRT_RFIscenario} for the P-band receiver we list values of the noise floor measured in two different days, when these RFI signals were (-180 $dBW m^{-2}Hz^{-1}$) or were not (-185 $dBW m^{-2}Hz^{-1}$) detected. Similarly for the L band, we report values for the noise floor of -174 and -172 $dBW m^{-2} Hz^{-1}$ measured in different days, showing the increase due to polluting emission when the antenna scan approaches the azimuth direction of the radar plant.
The last two columns of Table~\ref{SRT_RFIscenario} list the percentage RFI spectral occupancy calculated in the RAS and SR sub-bands, and in the receiver full frequency band respectively.
Preliminary measurements showed that, between the two measurement locations, the one closer to the telescope better represent what the telescope receivers detect in the same frequency range. 

Comparing the percentage spectral occupancy, the P band turns out to be the most polluted by RFI signal from power lines and digital links, with a slightly larger incidence in the horizontal polarization (43\%) than in the vertical one (38\%). Considering the RAS sub-bands in this frequency range, vertically polarized RFI signals were detected with spectral occupancy of 28\%. No horizontally polarized disturbances were measured in any of the RAS or SR frequency sub-bands. The RFI spectral occupancy calculation in the P band does not take into account the sporadic and unpredictable contribution of broad-band emission from power lines that, when detected, generally pollute most of the band making RFI monitoring almost useless and radio astronomical data very difficult to analyze. The presence of such disturbing signals has been reported to the national company in charge of the electric power distribution. Maintenance interventions in some of the power lines around the site have been executed, but further intervention would be needed to definitively solve the issue.
The L-band frequency range shows a prevalence of H-polarized RFI signals (24\%) with respect to V-polarized ones (12\%), mostly due to radar plants, cell phone networks and in a few cases to satellite down-links. In particular, impulsive emission from a radar plant pollute mainly two out of the five RAS frequency sub-bands. The 1330-1400 MHz RAS sub-band has a shared secondary use status and RFI occupancy percentages of 56\% and 11\% in H- and V-polarized signals. The 1400-1427 MHz RAS sub-band is allocated to exclusive primary use, no V-polarized disturbing signals were detected but the spectral occupancy due to the horizontally polarized ones reaches 44\%. RFI in the RAS sub-bands has been detected and characterized since the beginning of the monitoring with the mobile laboratory and the  unauthorized signals have been reported to the local and national authorities. 
The remaining frequency range up to 40 GHz appear to be essentially clean of RFI with worst-case occupancy percentages of the order of 8\%, except for some signals due to HiperLan and WiFi-LAN services in C-low and C-high bands as well as the effect of some digital links in the C-high, K, S, C-low and X frequency ranges. Remarkably, in the S frequency range the third RAS sub-band allocated to shared secondary use has spectral occupancy percentages of 28\% and 14\% in H- and V-polarization respectively. The X-frequency sub-band, allocated in shared primary use to the SR service, is scarcely polluted with RFI spectral occupancy percentages reaching 5\% and in H-polarization. It is worth noting that the K frequency range is almost clean except for the H-polarized disturbing signal from a broad-band digital link  occupying 5\% (about 30 MHz) of the total frequency bandwidth. 

\begin{figure}[h]
\centering
\includegraphics[width=38pc]{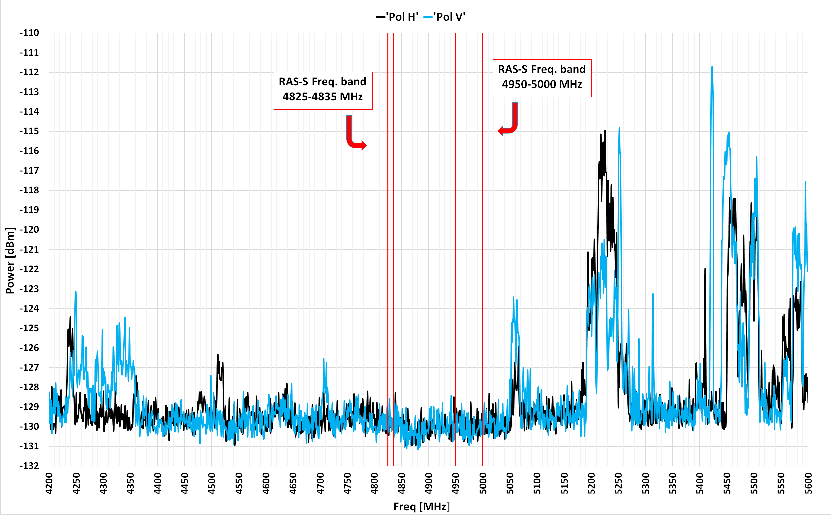}
\caption{C-low band spectra acquired during the 2016 RFI campaign at SRT. The horizontal (H-, black line) and the vertical (V-, light blue line) panoramic spectra were acquired at the location closer to the telescope with the spectrum analyzer in 3 MHz-RBW mode. The spectra power level has been corrected for the gain of the mobile laboratory receiver (channel D). The RAS secondary-allocation sub-bands are free from RFI, but the higher frequencies in the C-low range are getting increasingly polluted by broad-band radio services (HiperLan)}
\label{SRT_Low_Cband}
\end{figure}

Figure~\ref{SRT_Low_Cband} shows V- and H-polarized 3 MHz-RBW spectra in the band of the C-low receiver currently under construction, as measured during a dedicated RFI monitoring session. The RAS sub-bands 4.825-4.835 GHz and 4.95-5.0 GHz, shared as secondary use allocation with other services, appear clean of RFI but broad-band radio services (HiperLan) are increasingly occupying the higher portion of the C-low spectral range.

After the intensive RFI measurement campaign in 2016, the receiver frequency ranges of the SRT have been routinely monitored by means of both the fixed station, leveraging the same telescope sensitivity, and the mobile laboratory. A dedicated, real-time monitoring of the 100-2100 MHz IF baseband of the telescope, including the receiver polarizations, is currenlty available in piggy-back mode by means of a spectrum analyser and a dedicated LabVIEW software \cite{Serra2}. This allowed to update the information from the 2016 campaign, showing a little evolution of the RFI scenario at SRT with small differences mainly due to new RFI emitters.
Detrimental effects on radio astronomical experiments still persist in the P- and L-band frequency ranges. In addition, new RFI emissions at UHF frequencies caused by the enlargement of the national Terrestrial Trunked Radio (TETRA) network for emergency services and public safety started to affect the P band. Luckily, these disturbances are spatially confined in a narrow solid angle and at low antenna elevations ($\sim$ 6 deg). Also, the enlargement of the cell phone network with the installation of new 4G-LTE stations around the SRT site caused an increase of the RFI spectral occupancy in the L band above 1700 MHz.
In the C-low and C-high frequency ranges, in particular above 5000 MHz, the RFI spectral occupancy increased due to emission from new HiperLan services. Also, impulsive signals emitted by weather radar plant recently activated in the northern Sardinia have been detected. Also, the RFI monitoring in piggy-back mode allowed to more frequently detect emissions from different type of satellite services in the P, L and K bands.

\subsubsection{RFI measurement campaigns at the Noto telescope}\label{RFI Monitoring at Noto}
The 32-m Noto telescope is the Medicina twin antenna but, unlike the latter, it is equipped with an active surface for efficient high-frequency observations. The receiver frequency range (1.3-43.5 GHz) available for VLBI and single dish observations is summarized in Table~\ref{Noto_MobileStation}. 
Five dual-polarisation, mono-feed receivers are operating in the L, C-high, C-low, K and Q bands, the last three being cryogenically cooled. A coaxial, dual-polarization and dual-frequency receiver is available in the S/X bands.
It is worth noting that in Table~\ref{Noto_MobileStation} we consider the whole frequency range of the L-band receiver (1.3-1.8 GHz).
\\
Two first intensive RFI measurement campaigns, lasting one week each, were carried out in the frequency range 0.3-40 GHz at the Noto telescope in December 2014 \cite{Serra1} and May 2016  by means of the SRT RFI mobile laboratory, see Figure~\ref{MobileLabs}. Both campaigns were executed applying the {\it modus operandi} and RFI occupancy calculation described in Section~\ref{RFI Monitoring Sardinia-Sicily}, with the same spectrum analyzer settings. The comparison between homologous spectra acquired at different locations and in different epochs allows a characterisation of the RFI scenario and its evolution. A mountain plateau (Monte Rhenna) rising at 600 m above the sea level and located at about 3 km from the telescope was chosen as the measurement site for higher altitude monitoring. The second site, located close to the radio astronomical Noto Station, was chosen for RFI monitoring in the telescope proximity.

Table~\ref{Noto_RFIscenario} summarizes the results of the 2016 campaign in terms of percentage of the RFI spectral occupancy, computed mainly considering the measurements at the location closer to the telescope. The spectra acquired at the measurement site on Monte Rhenna were used only to better identify the RFI signals, but not to calculate the RFI spectral occupancy. The reader can refer to \cite{Serra1} for more details about the previous 2014 campaign. 
\\
For each frequency range listed in the first column of Table~\ref{Noto_RFIscenario} we report the noise floor measured in the 0.1 MHz- and 3 MHz-RBW spectra which, once expressed in term of $dBW m^{-2} Hz^{-1}$, has a similar value in both RBW configurations. Unlike the results of the RFI monitoring at SRT, the noise floor in the L band measured at the Noto radio astronomical Station was not affected by broad-band signals.
Comparing the L-band RFI spectral occupancy values in Tables ~\ref{Noto_RFIscenario} and ~\ref{SRT_RFIscenario}, it is evident that this frequency range is less polluted at Noto than at SRT (10\% against 24\%) irrespectively of the polarization. Although the RFI sources in this spectral range are the same (radar plants, satellites downlink and cell phone networks), the Noto radio astronomical Station is geographically less exposed. In fact, no signal was measured in the RAS sub-bands allocated for exclusive primary and shared secondary use, except for the 1330-1400 MHz one where the RFI spectral occupancy reaches 44\% (in both polarizations) due to radar emission.
Even if the telescope is not currently equipped with a P band receiver, during the 2016 campaign we monitored this frequency range also at Noto. Interestingly, detrimental effects similar to those observer at the SRT were measured in the P band due to power line emissions close to the Noto telescope. Those polluting signals were recorded in order to possibly report them to the national company responsible of the electric power distribution.

The S-band and C-low frequency ranges are almost clean except for some narrow-band signals that occupy a spectral percentage up to 8\% in vertical polarization in the S-band range. The remaining frequency ranges, C-high, X and K, resulted totally clear in the 2016 campaign. 

More recent measurements performed in 2021 with the new, portable Noto RFI station \cite{nicotra} confirmed the 2016 results for the C-high and X frequency ranges. However, the RFI scenario at the Noto telescope has been continuously evolving mainly in L, S and C-low bands due to radar plants emission and the increasing diffusion of signals coming from mobile phones, wi-fi and HiperLan networks.
Figure 36 in \cite{nicotra} shows the evolution of RFI signals in the receiver L-band at different epochs as measured with the SRT mobile laboratory and the new Noto RFI station.
The RFI spectral occupancy in the RAS L sub-bands has not changed, however it has significantly increased in the receiver full frequency range mostly due to radar and cell phone networks emission. In fact, L-band measurements done in 2021 nearby the Noto radio astronomical Station show an increased RFI spectral occupancy of 35\% with respect to the 10\% level measured in 2016, see Table~\ref{Noto_RFIscenario}). 
A significant increase of the RFI spectral occupancy in the S frequency range was measured as well, with percentages that in 2021 reached a value of 20\% in both polarizations due to Wi-Fi networks.

\begin{figure}[h]
\centering
\includegraphics[width=38pc]{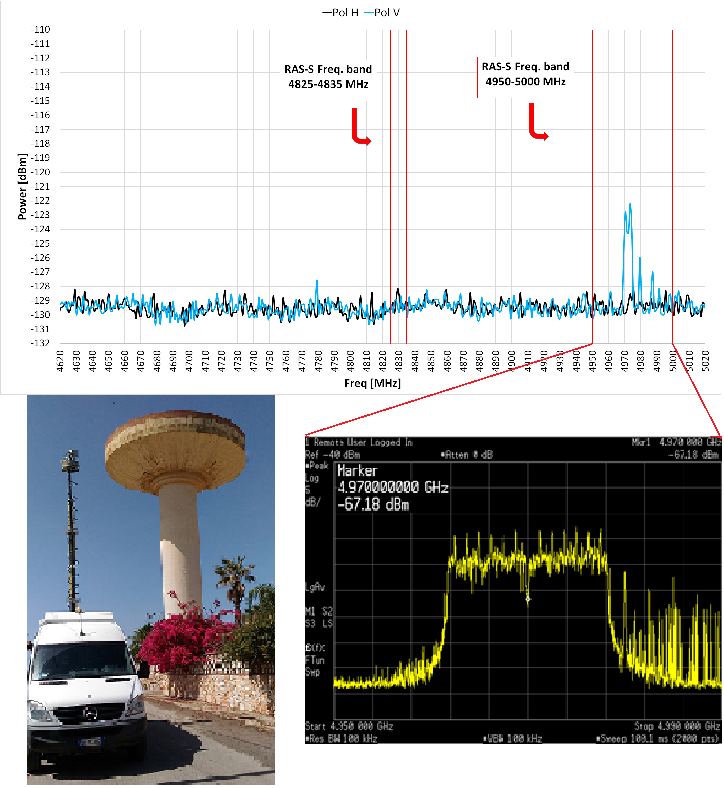}
\caption{C-low spectra acquired during the 2016 RFI monitoring campaign at Noto (top panel). The two H- (black line) and V- (light blue line) polarized spectra were acquired with a panoramic azimuth scan at the measurement site close to the telescope by setting the spectrum analyzer in 3 MHz-resolution bandwidth. The spectra power level has been compensated for the mobile laboratory receiver (channel D) gain. A zoom on the 4950-5000 MHz RAS secondary-use sub-band (bottom right panel) shows the broad-band (~16 MHz) HiperLAN V-polarized RFI and other related narrow-band polluting signals. Bottom left panel shows the mobile laboratory at the measurement site close to the water tank tower that hosted the HiperLAN antenna generating the RFI}
\label{Noto_Low_Cband_RAS_band2}
\end{figure}

For what concerns the C-low frequency range, some differences were measured in the 2016 spectra with respect to the 2014 and 2021 epochs. In fact, by comparing homologous spectra acquired at the location closer to the telescope, the frequency range was almost clear in 2014, with no polluting signal measured in the RAS secondary use sub-bands and only a few narrow-band RFI with spectral occupancy less than 1\% in the whole receiver frequency range. In 2016 the RFI spectral occupancy changed due to some V-polarized signals emitted by a Hiperlan network, occupying mainly the 4950-5000 MHz RAS secondary use sub-band. It is worth noting that HiperLANs are not included among the services sharing the use of this RAS sub-band. The broad-band ($\sim$16 MHz bandwidth) RFI emitted by an unauthorized HiperLAN transmitter has been detected in the frequency range 4962-4978 MHz mostly as a V-polarization signal, but a non-negligible H-polarization component has been measured as well (see Figure~\ref{Noto_Low_Cband_RAS_band2}). 
Follow-up investigations and signal maximization at three different measurement sites permitted to identify the polluting source, resulting to be a transmitting antenna mounted on top of a water tank tower located in a village about 7 km from the telescope (see bottom left panel in Figure~\ref{Noto_Low_Cband_RAS_band2}). The Noto RFI group reported the information on those interfering signals to the local administration deputed to enforcing the spectrum allocation. As a consequence, the HiperLAN transmitter was switched off, the C-low RAS sub-bands have been freed from RFI and the whole receiver frequency range was again only poorly polluted.  
Then, a recent radio monitoring performed in 2021 by using the new portable RFI station recorded new, unidentified broad-band H-polarized RFI signals in the frequency range 4920-4960 MHz. New investigations are needed to characterize this new HiperLan-like emission and to measure the consequent effective spectral occupancy.
\\
It is worth noting that measurement updates at higher frequencies in the K and Q bands will be possible with the extension of the spectral coverage of the portable RFI station, currently operating up to 18 GHz (\cite{nicotra}).

\begin{sidewaystable}
\caption{RFI scenario in the receivers band, RAS and SR sub-bands at the SRT. Sub-bands status: Exclusive Primary (ex-P), primary shared with other service (P), Secondary (S). No signal measured (NSM) = frequency ranges without RFI}
\label{SRT_RFIscenario}

\label{SRT_RFIoccup}
\centering
\renewcommand{\arraystretch}{0.94} 
\renewcommand{\tabcolsep}{0.01cm}
\begin{tabular}{l*{5}{c}}
\hline
\multirow{4}{8em}{Rx name \& freq range [GHz]} & \multirow{4}{8em}{RAS freq. range (Ex-P, P, S, SR) [GHz]} & \multirow{4}{10em}{M-NoiseFloor [$dBW m^{-2} Hz^{-1}$]} 
& \multirow{4}{9em}{RFI spect. Occ. w.r.t. RAS BW [\%] (Pol (H,V))} & \multirow{4}{9em}{RFI spect. Occ. w.r.t. Rx BW [\%] (Pol (H,V))}
\\
\\
\\
\\
\hline
\multirow{4}{6em} {P, .305-.410} & \multirow{4}{10em} {.322-.3286 (S)\hspace{2cm}  .400.6-401 (SR-P)  .406-.410 (P)} & \multirow{4}{10em}{-185 (-180)}  
& \multirow{4}{11em} {NSM (H)  28\% (V)\hspace{2cm}  NSM (H,V)\hspace{2cm}  NSM (H), 28\% (V)} & \multirow{4}{10em} {43\% (H), 38\% (V)}
\\
\\
\\
\\
\hline
\multirow{6}{6em} {L, 1.3-1.8} & \multirow{6}{10em} {1.330-1.400 (S)  1.400-1.427 (Ex-P)  1.6106-1.6138 (P)  1.660-1.670 (P)  1.7188-1.7222 (S)} & \multirow{6}{10em}{-174 (-172)} 
& \multirow{6}{11em} {56\% (H), 11\% (V)\hspace{2cm} 44\% (H), NSM(V)\hspace{2cm}  NSM (H,V)\hspace{2cm} NSM (H,V)\hspace{2cm} NSM (H,V)} & \multirow{6}{10em} {24\% (H), 12\% (V)}
\\
\\
\\
\\
\\
\\
\hline
\multirow{4}{8em} {S(UC) 3.0-4.5} & \multirow{4}{10em} {3.260-3.267 (S)  3.332-3.339 (S)  3.3458-3.3525 (S)} & \multirow{4}{10em}{-169} 
& \multirow{4}{11em} {NSM (H,V)\hspace{2cm}  NSM (H,V)\hspace{2cm}  28\%(H), 14\%(V)} & \multirow{4}{10em} {8\% (H), 2\% (V)}
\\
\\
\\
\\
\hline
\multirow{3}{10em} {C-low(UC), 4.2-5.6} &  \multirow{3}{10em} {4.825-4.835 (S), 4.95-5.0 (S)} & \multirow{3}{10em}{-166} 
& \multirow{3}{11em} {NSM (H,V)} & \multirow{3}{10em} {7\%(H), 5\%(V)}
\\
\\
\\
\hline
 \multirow{2}{8em} {C-high, 5.7-7.7} & \multirow{2}{10em} {6.650-6.6752 (S)} & \multirow{2}{10em}{-161} 
 & \multirow{2}{11em} {NSM (H,V)} & \multirow{2}{10em} {4\% (H), 4\%(V)}
\\
\\
\hline
\multirow{2}{8em} {X, 8.2-8.6} & \multirow{2}{10em} {8.4-8.5 (SR-P)} & \multirow{2}{10em}{-162} 
& \multirow{2}{11em} {5\%(H), 2\%(V)} & \multirow{2}{10em} {3\%(H), 3\%(V)}
\\
\\
\hline
 \multirow{7}{8em} {K, 18-26.5} & \multirow{7}{10em} {18.6-18.8 (SR-S), 21.2-21.4 (P), 22.0-22.5 (P), 22.81-22.86 (S), 23.6-24.0 (Ex-P), 25.5-27.0 (SR-P)} & \multirow{7}{10em}{-156} 
 & \multirow{7}{11em} {NSM (H,V)} & \multirow{7}{10em} {0.4\% (H), NSM(V)}
\\
\\
\\
\\
\\
\\
\\
 \hline
 \multirow{3}{6em} {Q(UC), 33-55} & \multirow{3}{10em} {36.43-36.5(S), 42.50-43.5(P), 48.94-49.04(P)} &\multirow{3}{10em}{-164} 
 & \multirow{3}{11em} {NSM (H,V)} & \multirow{3}{10em} {NSM(H,V)}

\\
\\ 
\end{tabular}
\end{sidewaystable}

\begin{sidewaystable}
\caption{RFI scenario in the receivers band, RAS and SR sub-bands at the Noto radio telescope. Sub-bands status: Exclusive Primary (ex-P), primary shared with other service (P), Secondary (S). No signal measured (NSM) = frequency ranges without RFI}
\label{Noto_RFIscenario}

\label{Noto_RFIoccup}
\centering
\renewcommand{\arraystretch}{0.8} 
\renewcommand{\tabcolsep}{0.1cm}
\begin{tabular}{l*{5}{c}}
\hline
\multirow{5}{8em}{Rx name \& freq range [GHz]} & \multirow{5}{9em}{RAS freq range (Ex-P, P, S, SR) [GHz]} & \multirow{5}{10em}{M-NoiseFloor [$dBW m^{-2} Hz^{-1}$]} 
& \multirow{5}{9em}{RFI spect. Occ. w.r.t. RAS BW [\%] (Pol (H,V))} & \multirow{5}{9em}{RFI spect. Occ. w.r.t. Rx BW [\%] (Pol. (H,V))}
\\
\\
\\
\\
\\
\hline
 \multirow{7}{8em} {L(UC), 1.3-1.8} & \multirow{7}{9em} {1.330-1.400 (S)  1.400-1.427 (Ex-P)  1.6106-1.6138 (P)  1.660-1.670 (P)  1.7188-1.7222 (S)} &\multirow{7}{9em}{-173} 
 & \multirow{7}{10em} {44\% (H), 44\% (V)\hspace{2cm} NSM (H) NSM(V)\hspace{2cm}  NSM (H,V)\hspace{2cm} NSM (H,V)\hspace{2cm} NSM (H,V)} & \multirow{7}{10em} {10\% (H), 10\% (V)}
\\
\\
\\
\\
\\
\\
\\
\hline
 \multirow{3}{8em} {S, 2.2-2.36} & \multirow{3}{12em} {No RAS band available} & \multirow{3}{9em}{-165} 
 & \multirow{3}{10em} {Not available} & \multirow{3}{10em} {3\% (H), 8\% (V)}
\\
\\
\\
\hline
 \multirow{3}{8em} {C-low, 4.62-5.02} &  \multirow{3}{9em} {4.825-4.835 (S) 4.95-5.0 (S)} & \multirow{3}{9em}{-165}  
 & \multirow{3}{7em} {NSM (H,V) NSM (H,V)} & \multirow{3}{9em} {\textless{1\%} (H), \textless{1\%} (V)}
\\
\\
\\
\hline
 \multirow{2}{8em} {C-high, 5.10-7.25} & \multirow{2}{9em} {6.650-6.6752(S)} & \multirow{2}{9em}{-161} 
 & \multirow{2}{10em} {NSM (H,V)} & \multirow{2}{10em} {NSM(H,V)}
\\
\\
\hline
 \multirow{2}{8em} {X, 8.18-8.58} & \multirow{2}{9em} {8.4-8.5(SR)} & \multirow{2}{9em}{-160} 
 & \multirow{2}{10em} {{NSM(H,V)}} & \multirow{2}{10em} {{NSM(H,V)}}
\\
\\
\hline
 \multirow{2}{8em} {K,21.5-23} & \multirow{2}{9em} {22.0-22.5(P), 22.81-22.86(S)} & \multirow{3}{9em}{-161} 
 & \multirow{2}{10em} {NSM (H,V)} & \multirow{2}{10em} {{NSM(H,V)}}
\\
\\
\\
\hline
 \multirow{2}{6em} {Q, 39.0-43.5} & \multirow{2}{9em} {36.43-36.5(S), 42.50-43.5(P)} & \multirow{2}{9em}{-166} 
 & \multirow{2}{10em} {NSM (H,V)} & \multirow{2}{10em} {NSM(H,V)}

\\ 
\end{tabular}
\end{sidewaystable}

\section{Mitigation strategies}

The characterization of the RFI local scenario is propedeutic to the implementation of RFI mitigation techniques based on on-line and off-line excision/cancellation to minimize the effect of RFI on the data. Such techniques include the use of digital spectrometers for data acquisition and the post-detection signal processing for RFI cleaning. 

Digital spectrometers characterised by high time and frequency sampling rate and broad-band are becoming the standard backends for radio astronomical observations. Field Programmable Gate Array (FPGA) technology are often used as the main computing engine and can be exploited to perform real-time RFI detection during data acquisition.
In recent years, state-of-the-art digital backends exploiting the publicly available software by the CASPER consortium (http://casper-dsp.org/) have been available at the SRT
and Medicina, and will be installed at Noto. 
In this evolving digital domain, telescope staff gained experience in the procurement and usage of FPGA-based systems as developed by the CASPER consortium, enabling the research of new solutions in the fields of beamforming and digital FX correlation \cite{stelio11} that can easily be applied to single dish data elaboration.
In order to improve our in-house capabilities in the design and implementation of digital spectrometers, we applied the experience gained in the use of such FPGA-based systems to the development of firmware tailored for the processing of single-dish data from the italian radio telescopes, as described in Section~\ref{spectrometer}.

Once spectropolarimetric data are acquired, RFI can be cleaned off-line in the spatial and spectral domain by means of dedicated software. To this aim we have developed a prototype tool for RFI inspection and flagging, described in Section~\ref{dishwasher}. This tool has been specifically designed to exploit the multi-feed capabilities of receivers installed at the Italian radio telescopes. DW is optimized for the flagging of huge amounts of data and can be fully integrated in a pipeline chain for single-dish data processing.

\subsection{The digital spectrometer}\label{spectrometer}

The hardware system provides analog to digital conversion, real-time elaboration in the digital domain and transmission of resulting data to the control system for further elaboration and storage. This is accomplished by means of a single ROACH board equipped with two iADC converter boards having 8 bit resolution and dynamical range of 48 dB and performing signal digitization at 2 Gigasample per second. The signal of one polarisation is converted to a digital stream of 8-bit signed data at 2 GHz. Data is then processed by means of the onboard XILINX Virtex5 FPGA device. 
With this hardware setup common to the three Italian radio telescopes, we developed a digital spectrometer delivering a bandwidth of 780 MHz for each polarization, with 1024 or 4096 frequency channels and 1 ms minimum dump time and integration period. The two polarizations are analyzed simultaneously and data are output over a 10 Gigabit ethernet link.So far, the digital spectrometer allows for total intensity measurements only.
To deal with high-power signals such as RFI, the firmware modules have been developed in order to optimize signal dynamics by monitoring possible overflows at each stage of the elaboration, starting from the digital conversion. Timing precision has been a design requirement in order to describe fast-moving or even transient phenomena with the necessary accuracy.

\begin{figure}[h]
\centering
\includegraphics[width=27pc]{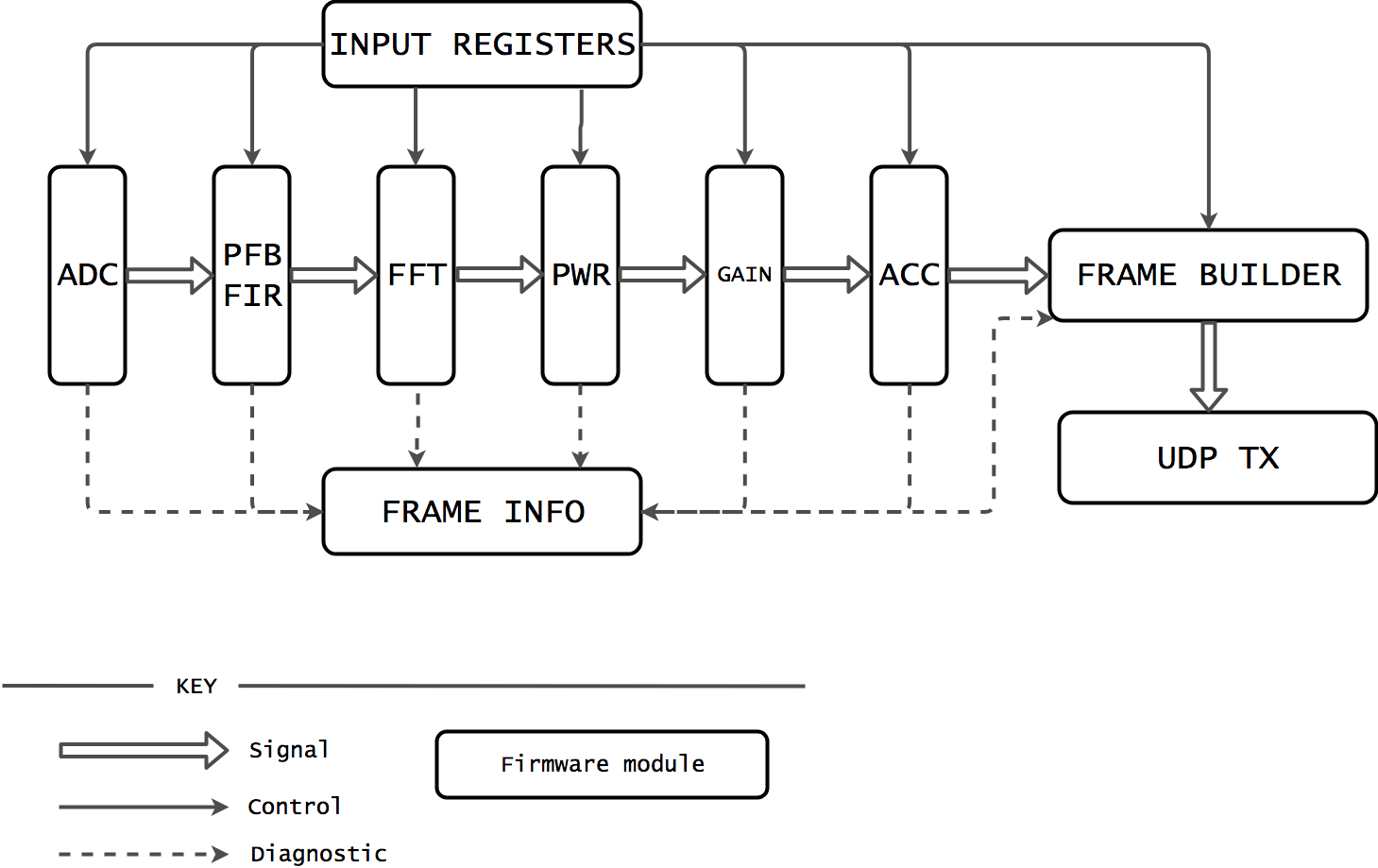}
\caption{Firmware elaboration stages of the digital spectrometer}
\label{firmware}
\end{figure}

\begin{figure}[!h]
\centering
\includegraphics[width=30pc]{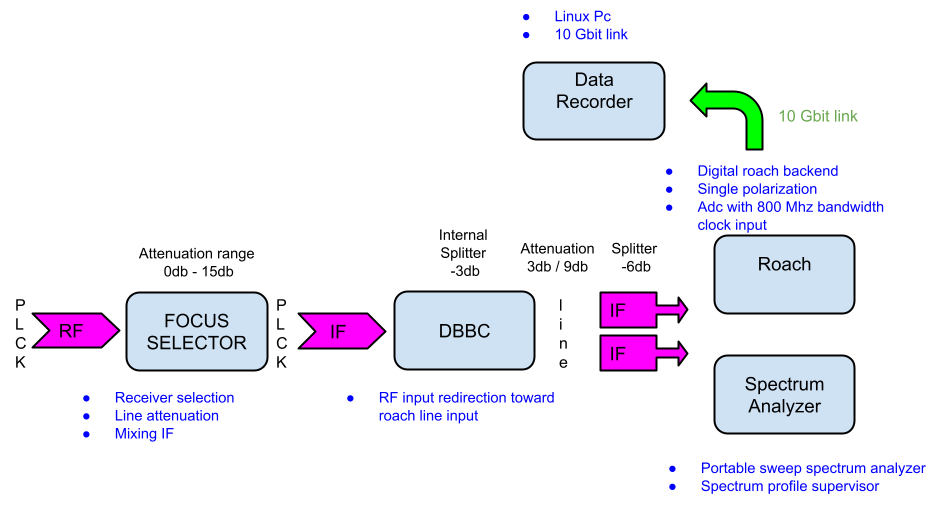}
\caption{Instrumental setup during test commissioning of the WBLGL spectrometer at the SRT}
\label{comm-setup}
\end{figure}

The digital chain follows two parallel paths for signal processing and for spectrometer control and metadata management. The firmware elaboration stages are described in Figure~\ref{firmware}. 
As data are converted into the digital domain by the ADC, a Polyphase Filter Bank (6 nodes, $>$50 dB channel isolation) transforms the time domain signal into the frequency domain outputting 4096 frequency bins in complex form.
The power spectrum is then computed and a configurable gain multiplies the obtained signal so that only the relevant bits are kept for further integration. 
Both the output data rate and the integration time can be controlled by integrating power spectra on board of the FPGA, exploiting memory cells on the FPGA itself.

\begin{figure}[h]
\centering
\includegraphics[width=33pc]{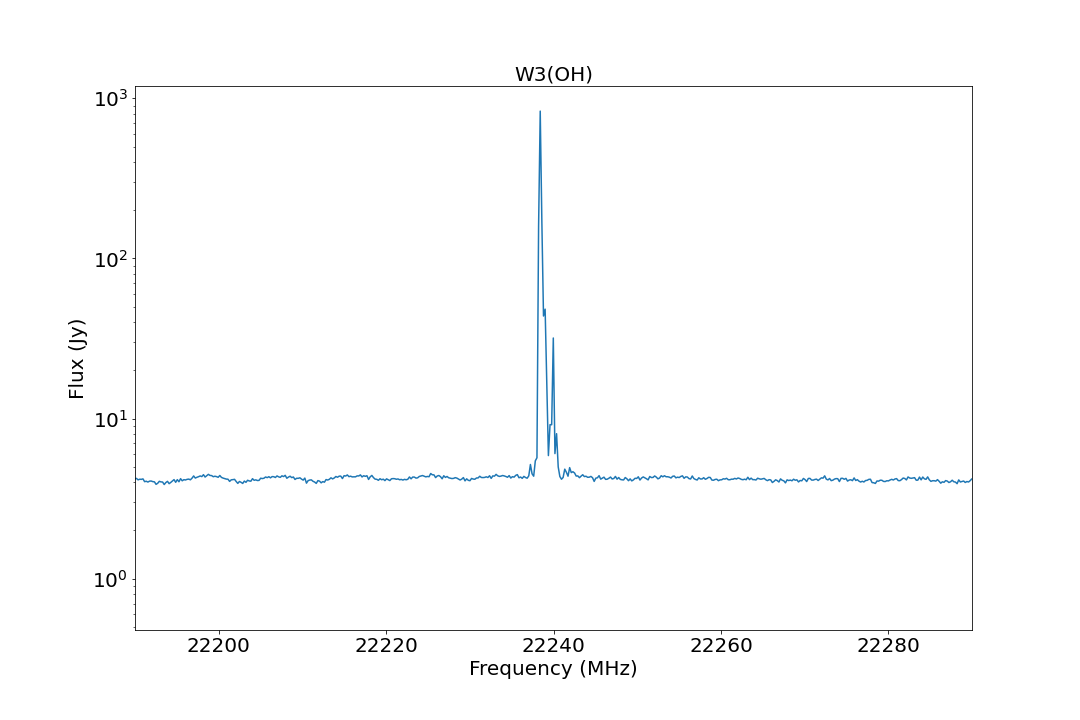}
\caption{ K-band emission of the W3(OH) water maser, acquired with position-switching observing mode and flux calibrated using the calibrator source 3C48}
\label{w3ohmaser}
\end{figure}

Control logics are implemented directly on the FPGA, where some registers are accessible to the control software via a PowerPC processor embedded into the ROACH board.
Data are time-stamped by means of a 64 bit posix \textit{TIME} register which sets the current time directly on the FPGA. Data acquisition and elaboration is continuous, the digital shift and gain can be tuned to the power level of the input signal and the number of power spectra to be accumulated in a frame is set by a dedicated register. Overflow diagnostic is executed at any digital processing stage to monitor the presence of artifacts created during the data elaboration in the presence of strong RFI signals. At the end of every integration, an UDP frame is created which also contains the meta information related to the integrated spectrum, like the gain and integration values, the overflow status and the posix timestamp. A basic Python-based control software module called WideBand Lowpass GigaBit (WBLGB) has been developed to access the ROACH functionalities. WBLGB makes use of the TCP/IP-based Karoo Array Telescope Control Protocol  \footnote{https://pythonhosted.org/katcp/index.html} to configure the board setup and control, like for instance the FPGA firmware version, the sampling frequency or network parameters for data transmission, and to inspect intermediate processing stages.

First observing tests aimed at validating the basic functionalities of the spectrometer were carried out at the Sardinia Radio Telescope. The observing setup is illustrated in Figure ~\ref{comm-setup}. The WBLGL backend delivered 4096 spectral channels over 780 MHz bandwidth on a single polarisation. Integrated spectra were continuously acquired with a time sampling of 2 ms, a limit set by the disk I/O performance of the recording system. To monitor the performance of the spectrometer, data acquisition was executed in parallel with both the ROACH backend and the spectrum analyser in use at the SRT.

\begin{sidewaysfigure}[p]
\centering
\includegraphics[width=1.0\textwidth]{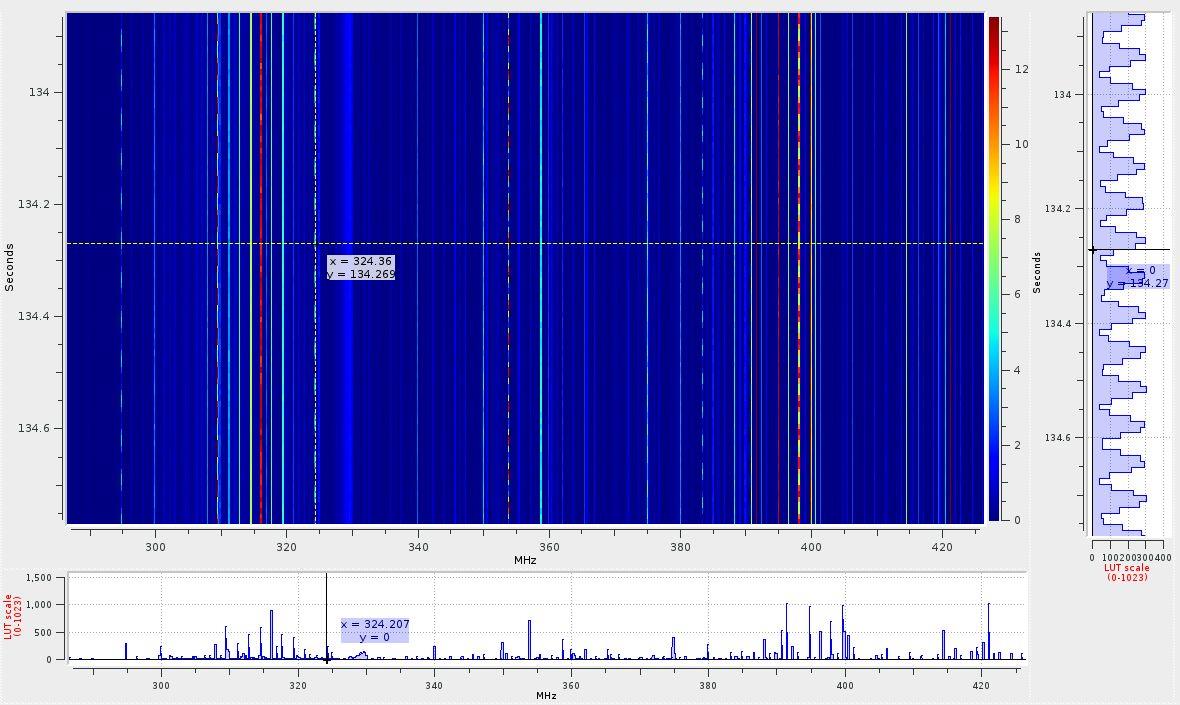}
\caption{Spectroscopic observation with the SRT P-band receiver, 10ms resolution, acquired with the digital spectrometer}
\label{specPband}
\end{sidewaysfigure}

The capability of WBLGL in detecting different types of RFI signal and its performances in terms of time and frequency resolution have been tested by observing well-known radio astronomical objects as well as some RFI sources known to affect the SRT site.
An estimate of the backend dynamic range has been obtained at 22 GHz, which is currently the frequency band less affected by RFI signals at the SRT (Figure ~\ref{w3ohmaser}).
Position-switching observations of the W3OH water maser were calibrated by means of the primary flux calibrator 3C48, whose flux is 1.2 Jy at 22~GHz. The backend dynamics allowed us to measure a line-to-noise ratio of $\approx$ 6000, in line with the expected performances of this type of digital instrumentation.

\begin{figure}[t]
\centering
\includegraphics[width=35pc]{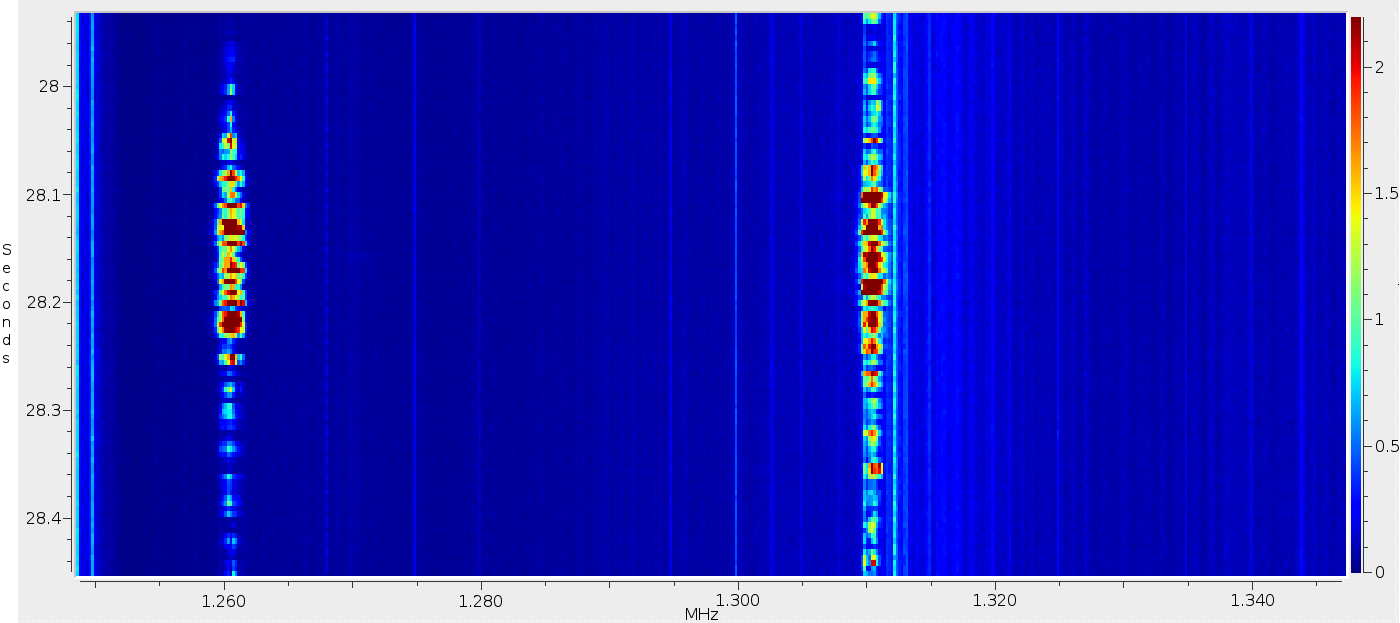}
\includegraphics[width=35pc]{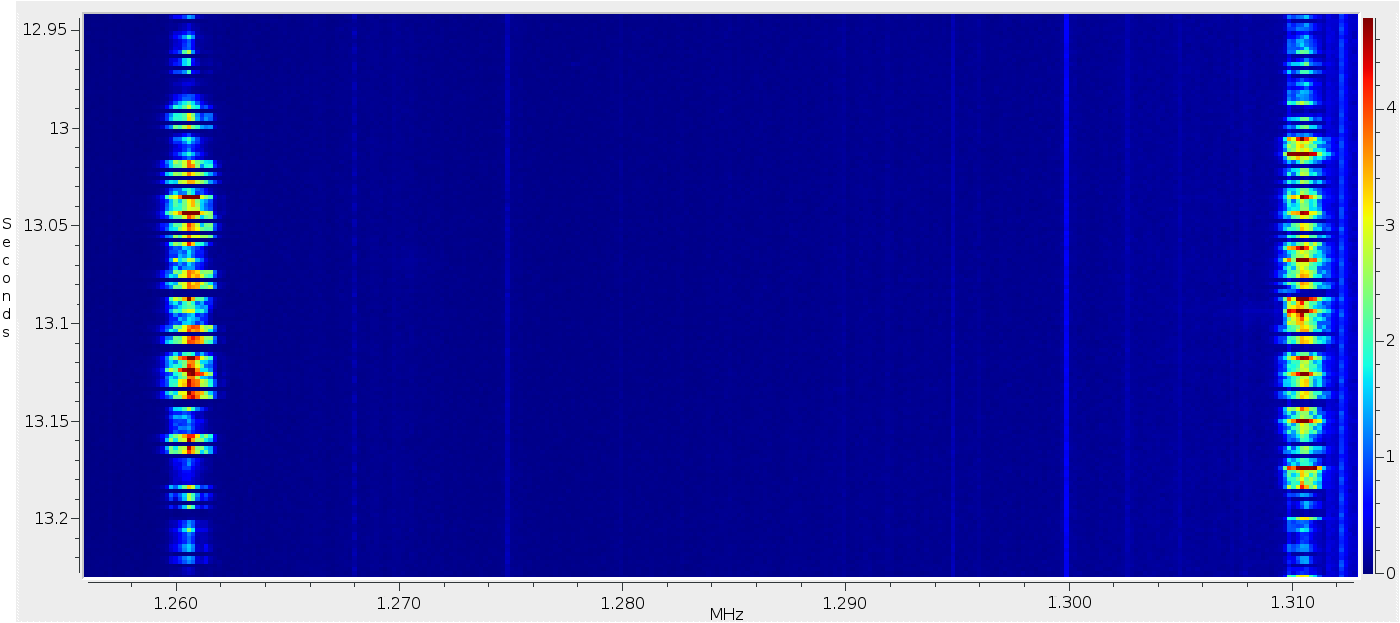}
\caption{SRT L-band spectrogram of a radar RFI signal detected by WBLGL with 5 ms (top) and 2 ms (bottom) resolution. Time intervals are 0.5 s and 0.3 s respectively}
\label{specLband}
\end{figure}

The spectrograms in Figure~\ref{specPband} and \ref{specLband} illustrate some examples of RFI signals measured in the P and L bands. The frequency on the horizontal axis refers to the RF of the signal chain, while the vertical axis shows the time from the start of the acquisition.

Figure~\ref{specPband} shows a 1~s-long acquisition with 10~ms temporal resolution in the P band with the telescope observing an azimuth scan at low elevation in the direction of a nearby power line. During this acquisition the well-known RFI caused by the power line (see Section \ref{RFI Monitoring at SRT}) was not detected, and as a consequence it was possible to detect a large number of other interfering signals. It was then possible to test the capability of the spectrometer in isolating the numerous RFI without overflowing. The 10~ms temporal resolution enables a detailed characterization of the terrestrial continuous signal, as can be seen in the sinusoidal view (vertical panel on the right of the plot) that describes the time domain evolution of the signal in the selected frequency bin.

Radars, satellite downlinks and cell phone network signals were detected when observing in the L band, as expected. In particular, an intermittent radar signal was observed to verify the capabilities in describing a short, periodic signal with different time resolutions. Acquisitions were done with time resolutions of 5 and 2 ms, this last being the maximum time resolution achievable with the adopted WBLGL configuration. 
Figure~\ref{specLband} illustrates the granularity with which the timing of the interfering signal can be described in a 0.5~s-long integration with 5 ms time resolution and in a 0.3~s-long integration with 2 ms time resolution, with negligible effects on the remaining spectrogram.

A strong RFI signal at 5896 MHz due to a digital link is continuously detected in the C band data and has been observed with different time resolutions of 10 and 100 milliseconds, and 1 second. No saturation issues were detected during the sampling and the subsequent digital elaboration stages.

Results of these tests demonstrated the robustness of WBLGL in detecting RFI signals with different characteristics in terms of periodicity, shape and in particular strength without incurring in instrumental saturation while preserving the measurement capabilities of the astronomical targets.

\subsection{Dish Washer: a prototype offline RFI Mitigation Tool}\label{dishwasher}

To complement the monitoring and mitigation activities described in the previous Sections we developed the Dish Washer (DW) interactive tool for single-dish data inspection, RFI detection and flagging \cite{zanichelli}. DW has been specifically built to handle the single-dish data delivered by the Italian radio telescopes (\cite{Righini15}) and is optimized for multi-feed receivers like those currently in use at Medicina and the SRT. Such receivers simultaneously output a large number of data streams from adjacent sky positions that may be affected by the same RFI signature and demand for a simultaneous processing for an efficient flagging.
As an example, the SARDARA backend \cite{melis18} at the SRT coupled with the 7-feeds K-band receiver may produce 7x4 full-Stokes spectropolarimetric data streams from a spatially coherent region of the sky, each stream consisting of up to 16384 spectral channels over a bandwidth up to 2 GHz. 

\begin{sidewaysfigure}[p]
\centering
\includegraphics[width=1.0\textwidth]{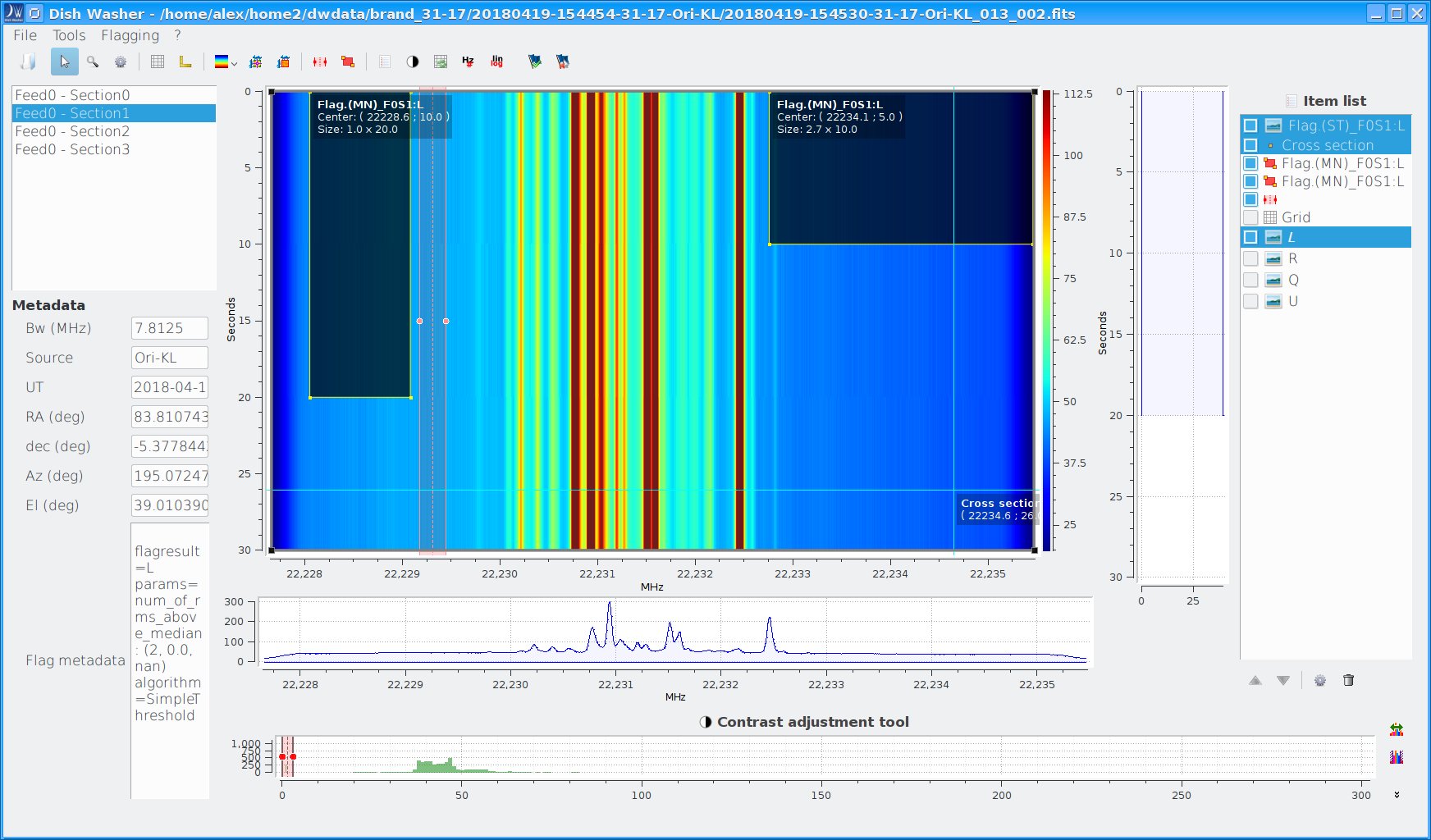}
\caption{Dish Washer GUI showing a flagging session for an observation of the star forming region Ori-KL. See text for a full description}
\label{dwgui}
\end{sidewaysfigure}

DW flagging can be performed by interactively inspecting the various data streams and the flagging process can be sped up by propagating the flagging table obtained for one data stream to the others, both in the spatial and the spectropolarimetric domain.
Flagged regions are listed in a dedicated FITS table associated to the input data and ready for use in subsequent data processing pipelines for actual RFI excision. Meta-data information is saved for each flagged region and method applied, to allow for the reconstruction of the flagging history for the given data set. An example of interactive DW session is shown in Figure \ref{dwgui}. 

From a software point of view DW consists of three Python sub-packages and a C library. The main sub-package implements core functionalities like data structure definition, I/O, manual flagging and the interface to the automatic flagging sub-package(s). Data formats other than those in use at the Italian radio telescopes can be supported by adding the proper methods of the relevant I/O class in a transparent way to the rest of the package.
A second sub-package implements the framework for the RFI detection algorithms and an interface to the C librabry which in turn implements flagging methods by taking advantage of C-language efficiency (including OpenMP parallel computing \footnote{http://www.openmp.org/}).
The third sub-package implements the graphical user interface (GUI) and acts as an interface to the main sub-package and its dependencies. Such an approach allows the functionalities provided by DW to be accessed in different ways and to be available in different scenarios. While the GUI is best suited for interactive flagging sessions, the main sub-package can be imported in a Python console of choice or a script in a batch processing scenario, or even integrated in a third-party application to provide data I\/O and flagging capabilities. Finally, the flagging Python sub-package and C library can be used to enable automatic flagging in applications with their own data format or different target use case.

Figure \ref{dwgui} shows an example of DW session. The leftmost area contains information on the dataset structure (receiver feeds and backend spectral sections) and the data subset to be flagged can be selected.  Meta-data relevant for RFI analysis such as the Horizontal coordinates, time and date of observation and flagging meta-data (if present) are also showed.
The central panel shows the dynamical spectrum of the selected subset, which can be inspected by means of cross cuts, zoom on areas, adjustments of contrast and intensity levels. Flagged areas are shown as shaded regions on the dynamical spectrum.
The rightmost part of the GUI list all the data streams of the selected subset - for instance the Stokes parameters - and the flagged regions that were created.
Buttons and menus at the top of the GUI enable interactive actions such as the identification and flagging of RFI-polluted regions, operations on the flagged regions and their propagation to other streams, e.g. to different polarization streams or to the several ones coming from the multiple feeds of a receiver array.

The first DW prototype has been released under GNU General Public License 3 and it is available as a Git repository at https://github.com/dishwasher-rfi . The current release includes the interactive graphical interface and very basic methods for manual flagging, based on the identification of RFI-polluted regions in the time and spectral domain. The implementation of automatic flagging procedures will be available in a future version of the DW code.

\section{Summary and Conclusions}\label{conclusions}

We presented a coordinated approach started at the INAF radio telescopes to deal with the increasing incidence of RFI. The project comprises three main activities: a fully systematic RFI monitoring at the observing sites, all year around, with dedicated receiving systems made of mainly commercially available instruments; the original development of digital backends, driven by firmware suited both for real time RFI identification as well as open to future online mitigation; finally, in the post-detection phase, the realization of an off-line tool for RFI flagging in the post-detection phase.

The coordinated RFI monitoring activity, mostly by means of mobile laboratory campaigns and with different approaches and methods applied at the three telescopes, has been illustrated. Results highlight the different RFI spectral occupancy of the same RF bands at the different sites, with the Medicina radio telescope being heavily polluted in the S and C frequency bands, where the diffusion of broad-band communication services is increasingly playing a crucial role, while the P band at SRT is heavily affected by power line-generated signals. The L frequency band (low plus high) is the one mostly polluted at each observing site, mainly due to interfering radar signals and mobile phone services. The RFI scenario is however undergoing a rapid evolution at all sites, as shown for instance by the increasing saturation of the C band at frequencies higher than 5000 MHz observed at Medicina and SRT. It is worth noting that the C band at Noto is still free from significant RFI but in some cases unauthorized emission has been reported in the 4950-5000 MHz RAS sub-band, which should thus be monitored with particular attention.
Mobile lab measurements provide an overview and a good estimation of the RFI spectral occupancy. They are best suited to identify specific terrestrial signal transmitters and to proceed with official reports to the local administrative authorities (aimed to switch them off) but cannot be considered as a complete characterization of the RFI all over the full RA frequency band of our receivers.
Continuous on-site monitoring through dedicated sessions for each receiver frequency range would be required, in order to leverage the synergies arising in putting together the different monitoring techniques exploited at the three sites according to their orography. Nevertheless, results of the monitoring campaigns provide a reference scenario useful to evaluate the evolution of the interference situation at the telescopes sites. 

The characterization of the RFI local scenario is very well suited to optimize the design of any following strategies of RFI mitigation techniques even if based on on-line or off-line excision/cancellation. 
We implemented a prototype digital spectrometer capable of high signal dynamics and precision timing to be used for on-line RFI mitigation. The ROACH-based hardware platform was equipped with a firmware in charge of signal processing, spectrometer control as well as management of observational data and metadata. The performances in terms of high dynamical range and sensitivity in both the frequency and time domains were successfully tested during an observing campaign at the SRT. The knowledge and expertise acquired in digital backends firmware development is propedeutic to future realization of backends capable to apply the complex techniques required by online RFI mitigation. Moreover, as the three Italian antennas share the same telescope operating system, the software implementation of the ROACH-based backend prototype and future, similar instrumentation can be easily exported from one radio telescope to another.

We are concentrating our efforts to apply efficient RFI flagging in the post-acquisition phase within the DW software tool, specifically designed for single-dish radio astronomical data. DW works off-line and is designed to be portable and flexible in order to easily accommodate future upgrades needed for counteracting new RFI environments.
The DW prototype has been tested on spectro-polarimetric observations from the Italian radio telescopes and is capable to visualize and manipulate the information from different feeds, spectral sections and polarizations contained in a single dataset. The possibility to propagate the flagging process to an entire data set (or among different data sets) allow an efficient processing of the huge data sets delivered by the modern digital backends. DW first prototype is currently released as ope-source software to the community and is available as a Git repository.

We believe that the described coordinated efforts were successful in identifying useful solutions to face the present RFI and limit its impact. They also pave the way to further developments, able to cope with the increasingly complex scenario and the new challenges we expect to encounter in a near future.
It is worth noting that a coordinated effort in RFI monitoring, together with the availability of similar observing equipment and control software at the Italian radio telescopes, has an immediate impact also on the coordination of the astronomical observations. This is particularly true for the two 32-m telescopes at Medicina and Noto, where observing programs can be in principle reallocated from one telescope to the other depending on the RFI spectral occupancy of the required receiver frequency band. This may also extend the concept of dynamical scheduling at the telescopes by adding a possible type of ``RFI forecast'' to the usual weather monitoring.
A very accurate knowledge of the RFI environment at the radio astronomical sites can also form the starting point for a national RFI archive. 
This would on one side facilitate the ongoing international effort for RFI monitoring and mitigation. On the other side, RFI surveys at the various sites would provide the astronomical community with valuable knowledge to plan their scientific observations in the most effective way.

\bigskip
\bigskip
{\bf Acknowledgements}
\noindent
This work has been conducted in the framework of the INAF PRIN 2012 Grant n. 2 ''RFI mitigation at the Italian radio telescopes'', P.I. Karl-Heinz Mack.

\noindent
We acknowledge J. Brand (INAF - Istituto di Radioastronomia) for having provided spectropolarimetric data to test the performances of the DW flagging tool.

\appendix
\appendixpage
\section{Main characteristics of the Medicina RFI fixed and mobile station receiving systems in the frequency bands of the radio astronomy receivers}

Tables~\ref{Medicina_FixedStationtable} and~\ref{MEDICINA_MObileStationTable} summarize the main characteristics of the receiving systems of the  fixed and mobile RFI stations, respectively, in the frequency bands of the radio astronomy receivers operating at the Medicina antennas \cite{Bolli3}.
The first column of both Tables~\ref{Medicina_FixedStationtable} and ~\ref{MEDICINA_MObileStationTable} lists the receivers frequency band, starting from the Northern Cross array one and continuing with those installed on the 32-m reflector antenna, including the under-construction (UC) cryogenic, dual-polarization and dual-feed Ku-band one.
RT and RX names refers to the radio telescope and the receiver names. In the second and third columns, we list the name, model, frequency band and isotropic gain of each antenna usually used to perform the RFI monitoring in the receiver frequency bands listed in the first column. In addition, for each of these frequency bands, the system gain (not including the antenna gain), the system noise figure and the system sensitivity evaluated at the physical temperature of 290 K and in a frequency bandwidth equal to 30 kHz are listed in fourth, fifth and sixth columns \cite{Ambros0, Ambros1, Ambros2, Ambros3}.    
 
\begin{sidewaystable}

\caption{Main characteristics of the fixed station receiving system for the RFI monitoring of the Medicina Radio telescopes frequency bands.}  
\label{Medicina_FixedStationtable}

\label{MEDICINA_MobileStationTable}

\centering
\renewcommand{\arraystretch}{0.9} 
\renewcommand{\tabcolsep}{0.1 cm}
\begin{tabular}{l*{6}{c}}
\hline
\multirow{2}{8em}{RT, Rx name, freq. range [GHz]} & \multirow{2}{8em}{Ant. mod., freq. band [GHz]} & \multirow{2}{10em}{Gain[dBi] @ [GHz]} & \multirow{2}{8em}{Channel Gain[dB] @[GHz]} & \multirow{2}{5em}{NF [dB] @[GHz]} & \multirow{2}{8em}{SysSen [dBW], (BW) [MHz]}
\\
\\
\\
 \hline
 \multirow{3}{8em} {N-Cross, P, .407-.409} & \multirow{3}{9em} {Yagi, .398-.418, LPA, 0.80-0.500} & \multirow{3}{6em} {13 @ .408, 10 @ .300}  & \multirow{3}{6em} {40 @ .408, 40 @ .325} & \multirow{3}{6em} {2.9 @ .408, 3.0 @ .325} & \multirow{3}{8em} {-156.5 (0.03), -156.4 (0.03)}
\\
\\
\\
 \hline
\multirow{3}{9em} {32-m refl., L, 1.35-1.45,1.595-1.715} & \multirow{3}{10em} {1.2 m-Reflec. ant., 1-12} & \multirow{3}{6em} {20-36 @ 1-12}  & \multirow{3}{7em} {30 @ 1.5} & \multirow{3}{6em} {5.3 @ 1.5} & \multirow{3}{8em} {-150.2 (0.03)}
\\
\\
\\

\hline
\multirow{3}{10 em} {32-m refl., S (coax), 2.20-2.36} & \multirow{3}{10em} {1.2 m-Reflec. ant., 1-12} & \multirow{3}{6em} {20-36 @ 1-12}  & \multirow{3}{7em} {26 @ 2.75} & \multirow{3}{6em} {6.4 @ 2.75} & \multirow{3}{8em} {-147.6 (0.03)}
\\
\\
\\
\hline
\multirow{2}{10em} {32-m refl., C-low, 4.30-5.80} & \multirow{2}{10em} {1.2 m-Reflec. ant., 1-12} & \multirow{2}{6em} {20-36 @ 1-12}  & \multirow{2}{7em} {21 @ 4.5} & \multirow{2}{6em} {8.9 @ 4.5} & \multirow{2}{8em} {-142.0 (0.03)}
\\
\\
\\
\hline
\multirow{3}{10em} {32-m refl., C-high, 5.90-7.10} & \multirow{3}{10em} {1.2 m-Reflec. ant., 1-12} & \multirow{3}{6em} {20-36 @ 1-12}  & \multirow{3}{7em} {43 @ 7.0} & \multirow{3}{6em} {6.9 @ 7.0} & \multirow{3}{8em} {-146.5 (0.03)}
\\
\\
\\
\hline
\multirow{3}{10em} {32-m refl., X (coax), 8.18-8.98} & \multirow{3}{10em} {1.2 m-Reflec. ant., 1-12} & \multirow{3}{6em} {20-36 @ 1-12}  & \multirow{3}{7em} {33 @ 11.0} & \multirow{3}{6em} {8.4 @ 11.0} & \multirow{3}{9em} {-143.2 (0.03)}

\\
\\
\\
\hline
\multirow{3}{8em} {32-m refl., K (2-beam), 18-26.5} & \multirow{3}{10em} {0.3 m-Reflec. Ant, 22-24} & \multirow{3}{6em} {35 @ 23}  & \multirow{3}{7em} {38 @ 22.5} & \multirow{3}{6em} {4.0 @ 22.5} & \multirow{3}{9em} {-153.5 (0.03)}

\end{tabular}
\end{sidewaystable}

\begin{sidewaystable}

\caption{Main characteristics of the mobile station receiving system for the RFI monitoring of the Medicina Radio telescopes frequency bands.}

\label{MEDICINA_MObileStationTable}

\centering
\renewcommand{\arraystretch}{0.9} 
\renewcommand{\tabcolsep}{0.1 cm}
\begin{tabular}{l*{6}{c}}
\hline
\multirow{3}{10em}{RT, Rx name, freq. range [GHz]} & \multirow{3}{8em}{Ant. mod., freq. band [GHz]} & \multirow{3}{10em}{Gain[dBi] @ [GHz]} & \multirow{3}{8em}{Channel Gain[dB] @[GHz]} & \multirow{3}{5em}{NF [dB] @[GHz]} & \multirow{3}{8em}{SysSen [dBW], (BW) [MHz]}
\\
\\
\\
 \hline
 \multirow{3}{8em} {N-Cross, P, .407-.409} & \multirow{3}{11em} {LPA, .310-.620} & \multirow{3}{5em} {5.5 @ .408}  & \multirow{3}{7em} {32.5 @ .408} & \multirow{3}{6em} {5.5 @ .408} & \multirow{3}{9em} {-149.6 (0.03)}
\\
\\
\\
 \hline
\multirow{3}{10 em} {32-m refl., L, 1.35-1.45,1.595-1.715} & \multirow{3}{10em} {DRGH, 1-18} & \multirow{3}{6em} {4 @ 1.7}  & \multirow{3}{7em} {32.5 @ 1.7} & \multirow{3}{6em} {3.5 @ 1.7} & \multirow{3}{9em} {-154.8 (0.03)}
\\
\\
\\
\hline
\multirow{3}{10 em} {32-m refl., S (coax), 2.20-2.36} & \multirow{3}{10em} {DRGH, 1-18} & \multirow{3}{6em} {4 @ 2.0}  & \multirow{3}{7em} {32.5 @ 2.0} & \multirow{3}{6em} {3.5 @ 2.0} & \multirow{3}{9em} {-154.8 (0.03)}
\\
\\
\\
\hline
\multirow{3}{10em} {32-m refl., C-low, 4.30-5.80} & \multirow{3}{10em} {DRGH, 1-18} & \multirow{3}{6em} {5 @ 5.0}  & \multirow{3}{7em} {30.5 @ 5.0} & \multirow{3}{6em} {3.0 @ 5.0} & \multirow{3}{9em} {-156.2 (0.03)}
\\
\\
\\
\hline
\multirow{3}{10em} {32-m refl., C-high, 5.90-7.10} & \multirow{3}{10em} {DRGH, 1-18} & \multirow{3}{6em} {5 @ 6.0}  & \multirow{3}{7em} {30.5 @ 6.0} & \multirow{3}{6em} {3.0 @ 6.0} & \multirow{3}{9em} {-156.2 (0.03)}
\\
\\
\\
\hline
\multirow{3}{10em} {32-m refl., X (coax), 8.18-8.98} & \multirow{3}{10em} {DRGH, 1-18} & \multirow{3}{6em} {6 @ 9.0}  & \multirow{3}{7em} {29 @ 9.0} & \multirow{3}{6em} {4.0 @ 9.0} & \multirow{3}{9em} {-153.4 (0.03)}
\\
\\
\\
\hline
\multirow{3}{10em} {32-m refl., ku (UC) (2-beam), 13.5-18} & \multirow{3}{8em} {DRGH, 1-18} & \multirow{3}{6em} {8 @ 16.0} & \multirow{3}{7em} {17 @ 16.0} & \multirow{3}{6em} {5.5 @ 16.0} & \multirow{3}{6em} {-149.6 (0.03)}
\\
\\
\\
\hline
\multirow{3}{8em} {32-m refl., k (2-beam), 18-26.5} & \multirow{3}{10em} {horn ant., 18-26.5, horn ant., 18-40} & \multirow{3}{5em} {20 @ 22.0, 14 @ 22.0}  & \multirow{3}{7em} {57 @ 22.0} & \multirow{3}{6em} {3.0 @ 22} & \multirow{3}{9em} {-156.2 (0.03)}
\\
\\
\\

\end{tabular}
\end{sidewaystable}

\section{Configurations and instrumentation settings of the SRT mobile receiving system during the RFI surveys at Sardinian and Sicilian
telescopes}
\label{RFIlab_conf_settings}

The configurations and the instrumentation characteristics of the mobile station receiving system, chosen to perform the radio surveys in the frequency range 0.3-40 GHz at SRT and Noto radiotelescope, are summarized in Tables~\ref{SRT_MobileStation} and \ref{Noto_MobileStation} respectively.
The first column of both tables list the name and the frequency range of the receivers working or under construction (UC) at the two radiotelescopes \cite{Bolli3}. In addition, for each of the frequency bands listed in the first column, the antenna and radio frequency channel characteristics are summarized from the second to the sixth column. The channel name and its frequency range (second column), the model, the isotropic gain and the half power beam width (HPBW) of the antenna installed on the top of the van mast (third column), the channel gain (not including the antenna gain, fourth column), the system noise figure (NF, fifth column) and the system sensitivity (last column) are listed. The system sensitivity has been calculated by adding the system NF (fifth column) to the thermal noise at 290 K evaluated in a 30 kHz frequency band. 

\begin{sidewaystable}

\caption{Configurations and instrumentation characteristics of the SRT mobile lab during RFI measurements at Sardinian radiotelescope}

\label{SRT_MobileStation}
\centering
\renewcommand{\arraystretch}{0.9} 
\renewcommand{\tabcolsep}{0.1 cm}
\begin{tabular}{l*{6}{c}}
\hline
\multirow{3}{8em}{Rx name, freq range [GHz]} & \multirow{3}{8em}{Channel name, freq range [GHz]} & \multirow{3}{10em}{Antenna \#, Gain[dBi], HPBW [deg] @[GHz]} & \multirow{3}{8em}{Channel Gain[dB] @[GHz]} & \multirow{3}{5em}{ NF [K] @[GHz]} & \multirow{3}{8em}{SysSen [dBW], (BW) [MHz]} 
\\
\\
\\
 \hline
\multirow{3}{8em} {P, .305-.410} & \multirow{3}{8em} {A, .30-.42} & \multirow{3}{8em} {LPA-370-10, 11, 48 @ .370}  & \multirow{3}{7em} {36 @ .370} & \multirow{3}{6em} {3.9 @ .370} & \multirow{3}{6em} {-153.7 (0.03)} 
\\
\\
\\
 \hline
\multirow{3}{8em} {L, 1.3-1.8} & \multirow{3}{8em} {B, 1.215-1.805} & \multirow{3}{8em} {LPA-2000-10, 11, 51 @ 1.5}  & \multirow{3}{7em} {30 @ 1.5} & \multirow{3}{6em} {4.0 @ 1.5} & \multirow{3}{6em} {-153.4 (0.03)} 
\\
\\
\\
\hline
\multirow{3}{8em} {S(UC), 3.0-4.5} & \multirow{3}{8em} {D, 3.3-5.5} & \multirow{3}{8em} {LPA 2000-10, 10, 60 @ 3.3}  & \multirow{3}{7em} {54 @ 3.3} & \multirow{3}{6em} {4.6 @ 3.3} & \multirow{3}{6em} {-151.9 (0.03)} 
\\
\\
\\
\hline
\multirow{3}{10em} {C-low(UC), 4.2-5.6} & \multirow{3}{8em} {D, 3.3-5.5} & \multirow{3}{8em} {DRGH 118-A, 11, 50  @ 4.9}  & \multirow{3}{7em} {52 @ 4.9} & \multirow{3}{6em} {5.7 @ 4.9} & \multirow{3}{6em} {-149.2 (0.03)} 
\\
\\
\\
\hline
\multirow{3}{8em} {C-high, 5.7-7.7} & \multirow{3}{8em} {E, 5.4-9.0 } & \multirow{3}{8em} {DRGH 118-A, 12, 48 @ 6.7}  & \multirow{3}{7em} {49 @ 6.7} & \multirow{3}{6em} {6.2 @ 6.7} & \multirow{3}{6em} {-148.0 (0.03)} 
\\
\\
\\
\hline
\multirow{3}{8em} {X, 8.2-8.6} & \multirow{3}{8em} {E, 5.4-9.0} & \multirow{3}{8em} {DRGH 118-A, 11, 49 @ 8.4}  & \multirow{3}{7em} {48 @ 8.4} & \multirow{3}{6em} {6.0 @ 8.4} & \multirow{3}{6em} {-148.5 (0.03)} 
\\
\\
\\
\hline
\multirow{3}{8em} {k, 18-26.5} & \multirow{3}{8em} {G, 18-40} & \multirow{3}{9em} {DRGH 1840-A, 13, 14 @ 22}  & \multirow{3}{7em} {68 @ 22} & \multirow{3}{6em} {3.8 @ 22} & \multirow{3}{6em} {-154.0 (0.03)} 
\\
\\
\\
\hline
\multirow{3}{8em} {Q(UC), 33-55} & \multirow{3}{8em} {G, 18-40} & \multirow{3}{9em} {DRGH 1840-A, 22, 7 @ 33}  & \multirow{3}{7em} {64 @ 33} & \multirow{3}{6em} {3.0 @ 33} & \multirow{3}{6em} {-156.2 (0.03)} 

\end{tabular}
\end{sidewaystable}

\begin{sidewaystable}

\caption{Configurations and instrumentation characteristics of the SRT mobile lab during RFI measurements at Noto telescope}

\label{Noto_MobileStation}
\centering
\renewcommand{\arraystretch}{0.9} 
\renewcommand{\tabcolsep}{0.1 cm}
\begin{tabular}{l*{6}{c}}
\hline
 \multirow{3}{8em}{Rx name, freq range [GHz]} & \multirow{3}{8em}{Channel name, freq range [GHz]} & \multirow{3}{10em}{Antenna \#, Gain[dBi], HPBW [deg] @[GHz]} & \multirow{3}{8em}{Channel Gain[dB] @[GHz]} & \multirow{3}{5em}{ NF [dB] @[GHz]} & \multirow{3}{8em}{SysSen [dBW], (BW) [MHz]} 

\\
\\
\\
\hline
\multirow{3}{8em} {L (UC), 1.3-1.8} & \multirow{3}{8em} {B, 1.215-1.805} & \multirow{3}{8em} {LPA-2000-10, 11, 51 @ 1.5}  & \multirow{3}{7em} {30 @ 1.5} & \multirow{3}{6em} {4.0 @ 1.5} & \multirow{3}{6em} {-153.4 (0.03)} 
\\
\\
\\
\hline
\multirow{3}{8em} {S, 2.2-2.36} & \multirow{3}{8em} {C, 2.185-3.288} & \multirow{3}{8em} {LPA 2000-10, 11, 57 @ 2.5}  & \multirow{3}{7em} {25 @ 2.5} & \multirow{3}{6em} {5.0 @ 2.5} & \multirow{3}{6em} {-150.9 (0.03)} 
\\
\\
\\
\hline
\multirow{3}{8em} {C-low, 4.62-5.02} & \multirow{3}{8em} {D, 3.3-5.5} & \multirow{3}{9em} {DRGH 118-A, 11, 50  @ 4.9}  & \multirow{3}{7em} {52 @ 4.9} & \multirow{3}{6em} {5.7 @ 4.9} & \multirow{3}{6em} {-149.2 (0.03)} 
\\
\\
\\
\hline
\multirow{3}{8em} {C-high, 5.10-7.25} & \multirow{3}{8em} {E, 5.4-9.0} & \multirow{3}{8em} {DRGH 118-A, 12, 48 @ 6.7}  & \multirow{3}{7em} {49 @ 6.7} & \multirow{3}{6em} {6.2 @ 6.7} & \multirow{3}{6em} {-148.0 (0.03)} 
\\
\\
\\
\hline
\multirow{3}{8em} {X, 8.18-8.58} & \multirow{3}{8em} {E, 5.4-9.0} & \multirow{3}{8em} {DRGH 118-A, 11, 49 @ 8.4}  & \multirow{3}{7em} {48 @ 8.4} & \multirow{3}{6em} {6.0 @ 8.4} & \multirow{3}{6em} {-148.5 (0.03)} 
\\
\\
\\
\hline
\multirow{3}{8em} {k,21.5-23.0} & \multirow{3}{8em} {G, 18-40} & \multirow{3}{8em} {DRGH 1840-A, 13, 14 @ 22}  & \multirow{3}{7em} {68 @ 22} & \multirow{3}{6em} {3.8 @ 22} & \multirow{3}{6em} {-154.0 (0.03)} 
\\
\\
\\
\hline
\multirow{3}{8em} {Q, 39.0-43.5} & \multirow{3}{8em} {G, 18-40} & \multirow{3}{8em} {DRGH 1840-A, 22, 7 @ 33}  & \multirow{3}{7em} {64 @ 33} & \multirow{3}{6em} {3.0 @ 33} & \multirow{3}{6em} {-156.2 (0.03)} 

\end{tabular}
\end{sidewaystable}

\end{document}